\documentclass{aa}  

\usepackage{graphicx}
\usepackage{txfonts}
\usepackage{natbib}
\usepackage[citecolor=blue,colorlinks=true,linkcolor=blue,urlcolor=blue]{hyperref}
\graphicspath{{./}{figures/}}
\usepackage[export]{adjustbox}
\usepackage{tablefootnote}

\begin{document} 

   \title{S 308 and other X-ray emitting bubbles around Wolf-Rayet stars}


\author{Francesco Camilloni 
    \inst{1}\thanks{\email{fcam@mpe.mpg.de}}
    \and Werner Becker \inst{1}\inst{,2}
    \and Manami Sasaki \inst{3}
}

\institute{Max-Planck-Institut für extraterrestrische Physik, Giessenbachstraße, 85748 Garching, Germany    
 \and
Max-Planck-Institut für Radioastronomie, Auf dem Hügel 69, 53121 Bonn, Germany
\and
 Dr. Karl Remeis Observatory, Erlangen Centre for Astroparticle Physics, Friedrich-Alexander-Universität Erlangen-Nürnberg,
Sternwartstrasse 7, 96049 Bamberg, Germany
}
   \date{Received 15 September 2023 / Accepted 1 November 2023}

 
  \abstract{S 308 is an X-ray emitting bubble that surrounds the Wolf-Rayet star WR6. The structure shines in the optical as well and is thus known as the Dolphin Nebula. Due to its large angular extent, it has been covered at only 90\% with past \textit{XMM-Newton} observations.
  }
  {Thanks to the unique dataset provided by the all-sky survey performed in  X-rays by \textit{SRG}/eROSITA, we can show for the first time the image of the bubble in its entire extent in this band, together with its spectral characterization. Moreover, we have tried to apply the same procedure for other  wind-blown bubbles detected in the optical/IR  and we searched for X-ray extended emission around them.
  }
  {We first analyzed the diffuse emission of S308, providing a detailed spectral analysis. We then considered a sample of 22 optical/IR selected wind-blown bubbles from a previous study based on \textit{WISE} data, providing an estimate of the X-ray flux for the first time. 
  }
  {We obtained the best fit for S308 with a two-temperature non-equilibrium plasma model (kT$_{1}=0.8_{-0.3}^{+0.8}$ keV and kT$_{2}=2_{-1}^{+3}$ keV) showing super-solar N abundance and low absorption. We did not detect any of the 22 optical/IR emitting bubbles in X-rays, but using our best fit model, we estimated the 3$\sigma$ flux upper limits for each bubble. 
  }
  {We demonstrate the new possibility offered by \textit{SRG}/eROSITA to study known wind-blown bubbles and look for other ones. A two-temperature  plasma description seems to fit the data quite well for S308. Since all of the 22 bubbles studied still remain undetected by \textit{SRG}/eROSITA, it is very likely that absorption effects and spatial compactness are responsible for the challenges standing in the way of detecting these bubbles in soft X-rays.}
\authorrunning{F.Camilloni, W. Becker \& M. Sasaki}
\titlerunning{S 308 and other X-ray shining bubbles}

   \maketitle
%

\section{Introduction \label{sec:intro}}
Wolf-Rayet (WR) stars are massive stars, responsible for carrying considerable amounts of material in the interstellar medium (ISM) via strong winds. Among the different phases of a massive star's life, the WR phase is the very late stage when the outer layers have  already been expelled and what is shining is just a very hot core. There are different types of WR stars, depending on which elements dominate their optical spectra. The main types are WN, WC, and WO, which have their spectra  dominated by nitrogen, carbon, and oxygen, respectively. Some WN stars show hydrogen in their spectra, while WO and WC do not show hydrogen. For a modern overview, see for example \cite{Sander2019}. As a consequence of this very powerful activity and expulsion of material from their envelope, WR stars play a key role in determining the composition of the ISM. Some authors even consider the possibility that the Solar System originated in the wind-blown bubble of a WR star \citep{Dwarkadas2017}. As a consequence, WR stars also play a key role in the star formation process.

Since WR stars are known to be in the very late stage of their lives, they are considered to be the closest example of stars that are going to explode as core-collapse supernovae \citep{Woosley2002,Sander2020} in a relatively short amount of time. This makes studies of these objects particularly interesting, not just for stellar physics, but also relating to our understanding of supernova explosion mechanisms and supernova remnant studies. At the time of writing (summer 2023), the Wolf-Rayet Star Catalogue \footnote{\url{https://pacrowther.staff.shef.ac.uk/WRcat/}} \citep{Rosslowe2015} lists 667 WR stars.

WR6 is a prototypical WR star which is visible as a naked eye object in the Canis Major constellation (also known as EZ Canis Maior, HD50896 \cite{vanderHucht2001}). Given it is very luminous also in the X-ray band, it has been covered by deep grating observations with \textit{Chandra} and \textit{XMM-Newton}. Making use of this rich dataset, \cite{Oskinova2012} and \cite{Huenemoerder2015} showed the presence of an intriguing plasma emitting structure. While most of the spectrum is well described by collisional equilibrium thermal plasma components, the very high quality of the spectra shows another component in the hard X-rays, whose origin is still unknown. It is well known that WR stars are surrounded by dense material, so it is possible that this additional component originates there. One noticeable example of hard X-ray emitting plasma around massive stars is found in colliding-wind binaries (CWB). However, works such as \cite{Pradhan2021} and \cite{Abaroa2023} demonstrate the description of the dense wind around WR stars is still far from being complete. On top of this unclear theoretical scenario, \cite{Koenigsberger2020} provided evidence for an unseen companion around WR6. This clearly could alter the interpretation given by the earlier studies of \cite{Oskinova2012} and \cite{Huenemoerder2015}, with the additional cold plasma component possibly due to reprocessed material around or by the unseen companion. In the context of having a possible double system in WR6, we notice how \cite{Apep2023} highlighted the presence of a hard X-ray emitting component around the so called \textit{Apep} system, which is composed of two WR stars.

Moreover, the WR6 system is particularly interesting also because it is surrounded by an extended circumstellar bubble called S308. Detected for the first time with \textit{ROSAT} \citep{Wrigge1999}, this structure was extensively studied in X-rays, employing several \textit{XMM-Newton} observations \citep{Chu2003,Toala2012}.  Given the large extent of the remnant, this dataset was barely sufficient to cover 90\% of its extent in  X-rays. As a consequence, also due to its very low surface brightness, this circumstellar bubble has never been imaged in its entire extent in the X-rays. In addition, the impossibility of finding a background that is fully independent from the source in the \textit{XMM-Newton} dataset led \cite{Toala2012} to conclude that their spectral analysis was likely to be biased for this reason.

Therefore, the launch of the X-ray telescope eROSITA in 2019 \citep{Predehl2021} on board the Spectrum-Roentgen-Gamma Observatory (\textit{SRG}, \cite{Sunyaev2021}) has provided a unique opportunity, thanks to its large field of view and \textit{XMM-Newton}-like spectral resolution. These factors provide the ideal conditions for studying extended sources, such as galaxy clusters, supernova remnants, and bubbles. Given the unique dataset provided by the first four all-sky survey scans of eROSITA (with the merged dataset called eRASS:4), in this paper, we release  the image and spectral analysis of the bubble in its entire extent, for the first time. Moreover, the eRASS:4 dataset allows us to look for X-ray emission around  22 similar bubbles that are known from their optical/IR emission \citep{Toala2015}. In Section \ref{sec:Data_reduction}, we describe the data reduction process, while in Section \ref{sec:Results}, we present the image and the spectral analysis carried out on S308 together with the flux estimate on the others bubbles. Finally, in Section \ref{sec:Discussion}, we discuss our results and in Section \ref{sec:Conclusion}, we present  our conclusions.

\section{Data reduction \label{sec:Data_reduction}}
\begin{figure*}
    \centering
    \includegraphics[scale=0.6]{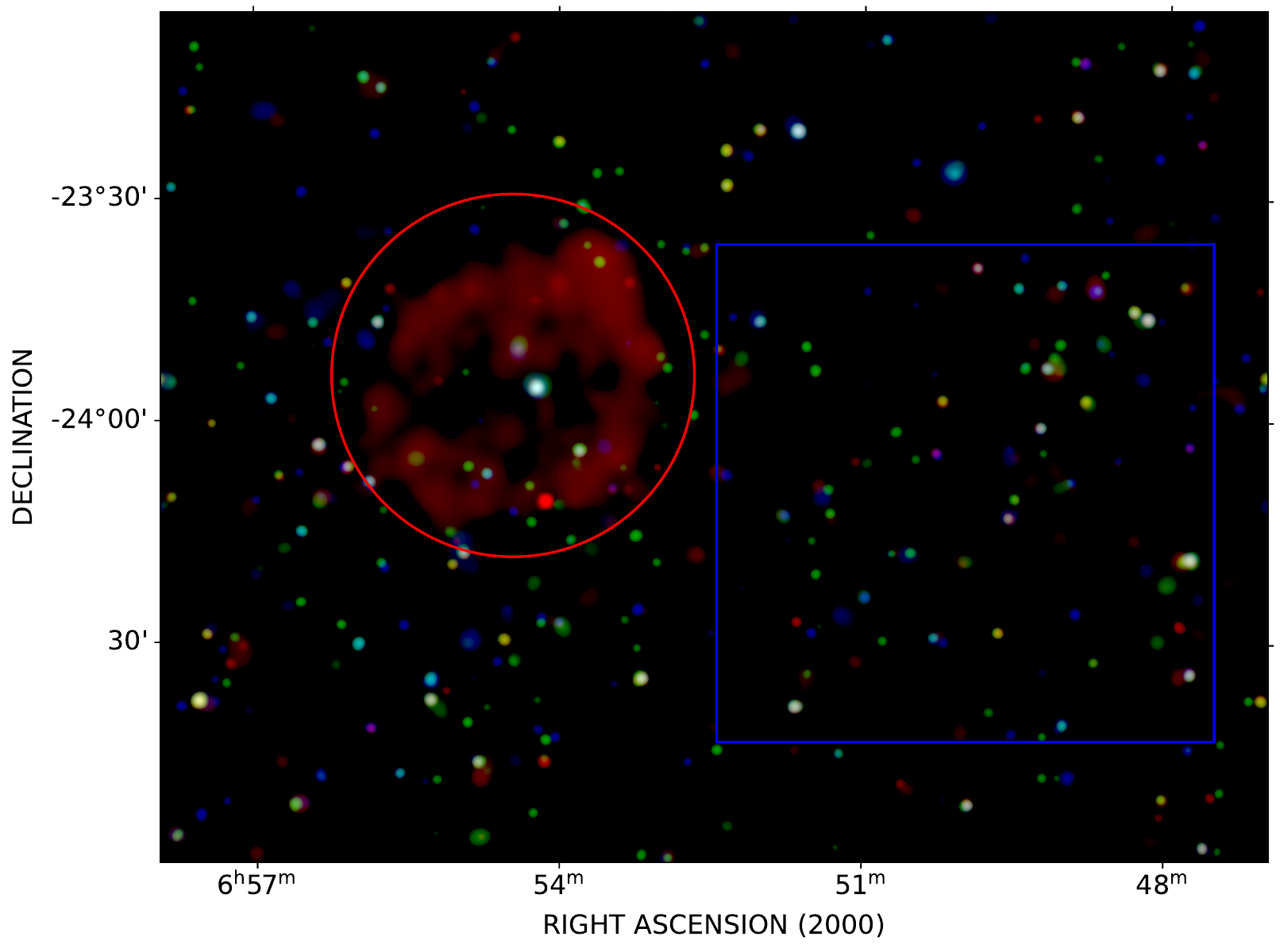}
    \caption{False color image (Red: 0.2-0.7 keV, Green: 0.7-1.1 keV, Blue: 1.1-10 keV) of the X-ray emitting bubble S308. The star WR6 shines in the center of the bubble. The red line indicates the source extraction region, while the blue line indicates the background one. The smoothing algorithm is based on the work of \cite{Ebeling2006} and is applied as described in Section \ref{sec:SpatialAnalysisS308}. The image scaling has been considerably stretched to highlight the faint soft X-ray diffuse emission of the nebula.
    \label{fig:S308_WR6}}
\end{figure*}

We started the data analysis with the reduction of the eROSITA dataset that passed over S 308 for four consecutive all-sky scans (April 11-19, 2020, October 15-22, 2020, April 10-21, 2021, and October, 12-27, 2021) summing up to a total observing time of 1860s. To carry out this operation, we employed the eROSITA science analysis software (eSASS) version 211214\footnote{Software version named after the release date: 14.12.2021} \citep{Brunner2022}. With the command \texttt{evtool}, we merged the event files from tiles 102114 and 105114, using the events from all the four scans (dataset called 'eRASS:4'). With the same command, we cleaned the merged event file from solar flaring, selecting all photon patterns available (PATTERN=15) in order to maximize the signal. To produce the image shown in Figure \ref{fig:S308_WR6}, we corrected for the vignetting over the whole field of view in the 0.2-10 keV band, obtaining a final vignetted-corrected exposure time of 586s.

For the spectral analysis, we analyzed the entire nebula, masking the point sources. After this masking operation, we finally extracted the spectrum, background, redistribution matrix file (RMF), and ancillary response file (ARF) with the command \texttt{srctool}. 

\begin{figure*}
    \centering
    \includegraphics[scale=0.4]{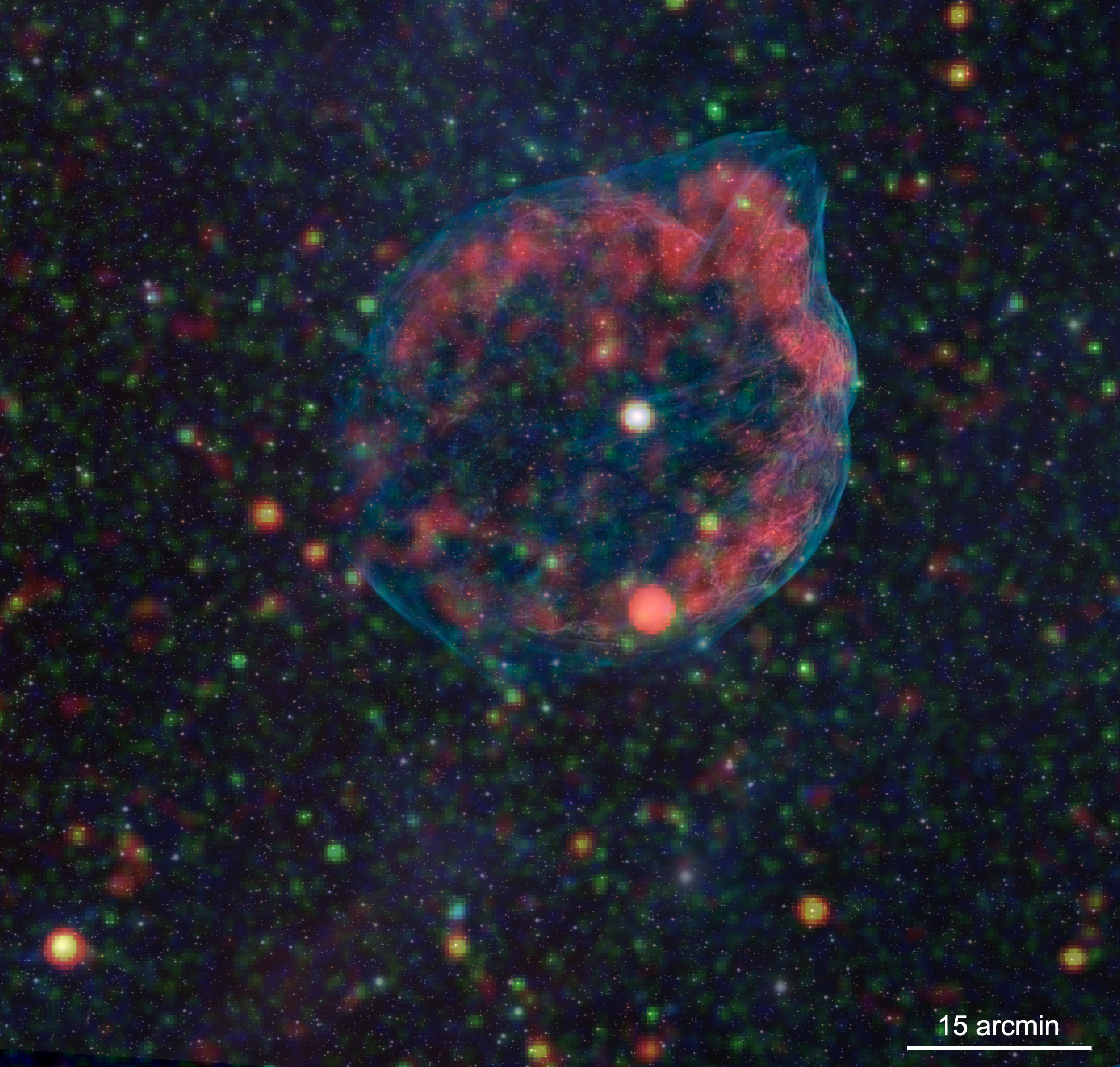}
    \caption{Composite eRASS:4 RGB and optical image of the S308 nebula.The eROSITA image was smoothed using a Gaussian filter with 1 $\sigma$ standard deviation value to calculate the Gaussian kernel. The colors in the X-ray image correspond to  R: 0.2-0.7 keV, G: 0.7-1.2 keV and B:1.2-8.0 keV. The optical image was overplotted onto the eRASS:4 image with a 50\% transparency. It was taken using an ASA 50cm reflecting telescope and H$\alpha$ and O[III]  filters. Optical image courtesy of Nik Szymanek \& Telescope Live. \label{fig:S308_optical_composite}}
\end{figure*}

\section{Results \label{sec:Results}}
\subsection{Spatial Analysis of S308 \label{sec:SpatialAnalysisS308}}

To analyze the diffuse emission from the X-ray emitting bubble, we started by creating the image shown in Figure \ref{fig:S308_WR6}. The image was smoothed with CIAO X-ray Data Analysis software \citep{CIAO} tool \texttt{csmooth}, using as the parameters a minimum significance of S/N=2 and maximum of 7, along with a minimum smoothing scale of  5 pixels and maximum smoothing scale of 20 pixels. As can be seen from Figure \ref{fig:S308_WR6}, the inner part of the bubble seems to be considerable less luminous than the outer parts. The straightforward justification for this would be that the intense winds coming from the bright central Wolf-Rayet star have swept out the material expelled from the star itself in a previous phase. We also created an optical/X-ray composite image as shown in Figure \ref{fig:S308_optical_composite}.

\subsection{Spectral analysis of S308 \label{sec:SpectralAnalysisS308}}
 To perform the spectral fitting analysis, we employed the Python interface of XSPEC \citep{Xspec}, called PyXSPEC, using the Cash statistics \citep{Cash1979} version implemented in XSPEC \citep[see][for a discussion]{Kaastra2017}. We will refer to this quantity later as CSTAT. We fitted the spectra from 0.3 keV to 10 keV. The errors on the parameters are expressed in 1 $\sigma$ confidence level (68\%). We extracted the spectra from the entire nebula and from the background using the regions shown in Figure \ref{fig:S308_WR6}. 

After removing the point sources, we started to characterize the background following the modeling described in \cite{Okon2021}. We employed three different spectral components: one power law to describe the cosmic X-ray background (CXB) with the slope fixed at 1.4 plus three thermal components (VAPEC, \cite{Smith2001}) representing the galactic halo and the Local Hot Bubble (LHB). The galactic halo is modeled by one component with kT$_{1}$=0.658 keV and a second component with kT$_{2}$=1.22 keV. The third component representing the LHB is modeled with a thermal model of temperature kT=0.105 keV. The normalization of all these four models are left free to vary. Before fitting this model to the background spectrum, we modeled the instrumental background using the best-fit model obtained from the filter wheel closed data, as described in \cite{Yeung2023}. In this way, we obtained the best-fit model for the background, accounting for both sky and instrumental contributions.

We then fit  the best-fit background model and the source model together to the source spectrum. As presented in \cite{Camilloni2023}, we took into account the geometric area of the extraction region (expressed in sr) in our spectral modeling ("constant" factor in front of the source model), which was kept fixed.  While fitting simultaneously source and background model, we kept  all the parameters frozen, except an overall global normalization in the background model. To better estimate the errors in the spectral fitting analysis, we used a code based on the Markov chain Monte Carlo  (MCMC) library \texttt{emcee} \citep{emcee2013}. We initialized our walkers using a Gaussian distribution centered on the parameters of a first fit run. We then explored the posterior using logarithmically uniform priors. We left the code run for 40000 steps, selecting only the last 2000 steps to ensure most of the walkers were converged. All the abundances are expressed in solar units.

\begin{figure*}
    \centering
    \includegraphics[scale=0.4]{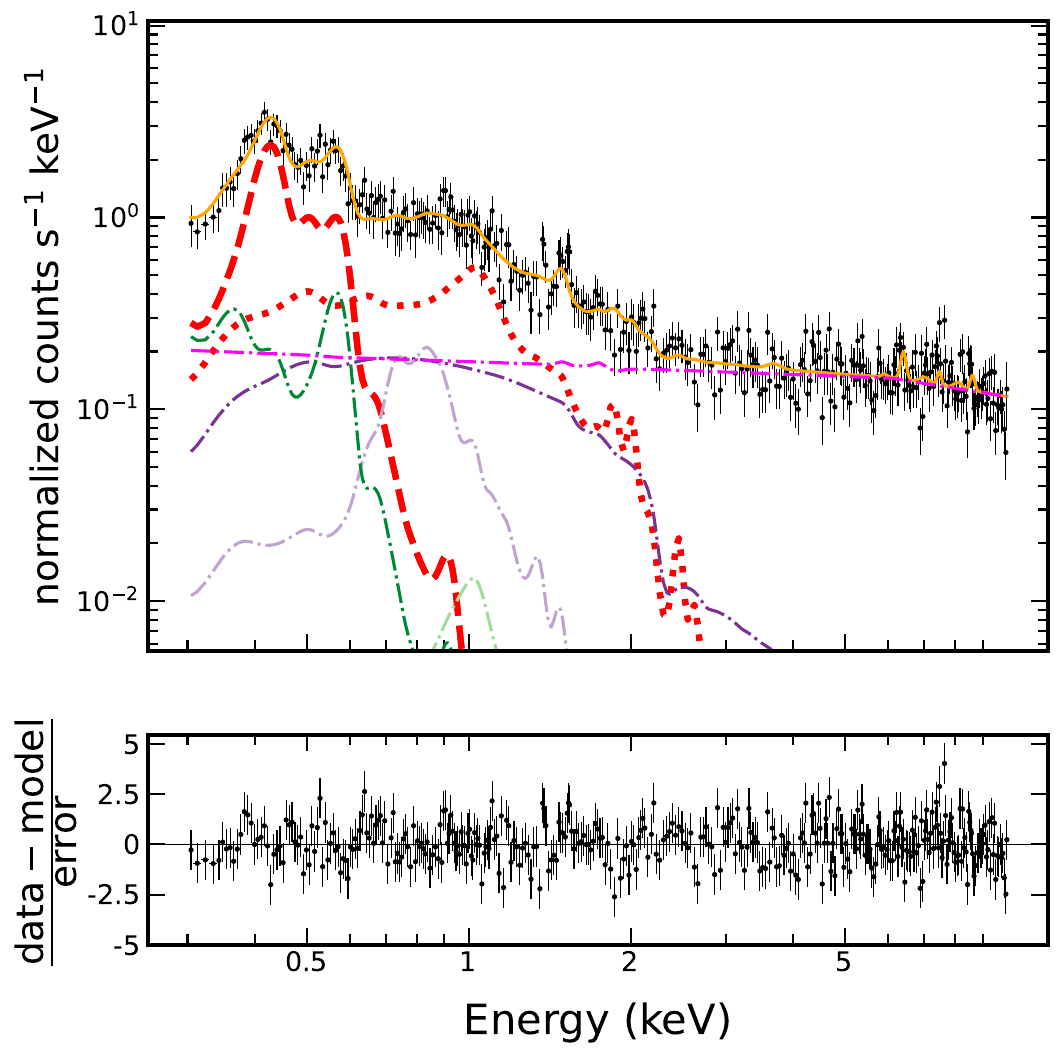}
    \includegraphics[scale=0.4]{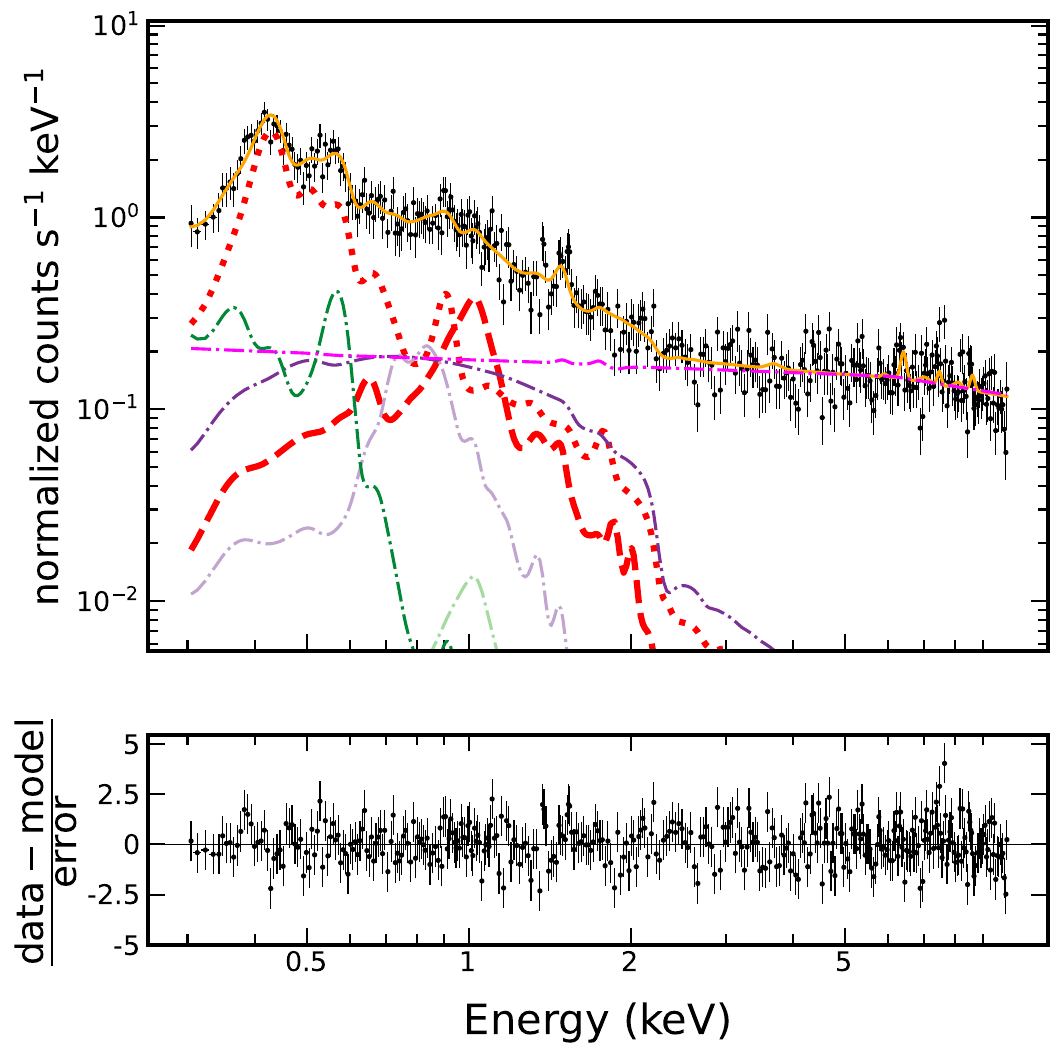}
    \includegraphics[scale=0.4]{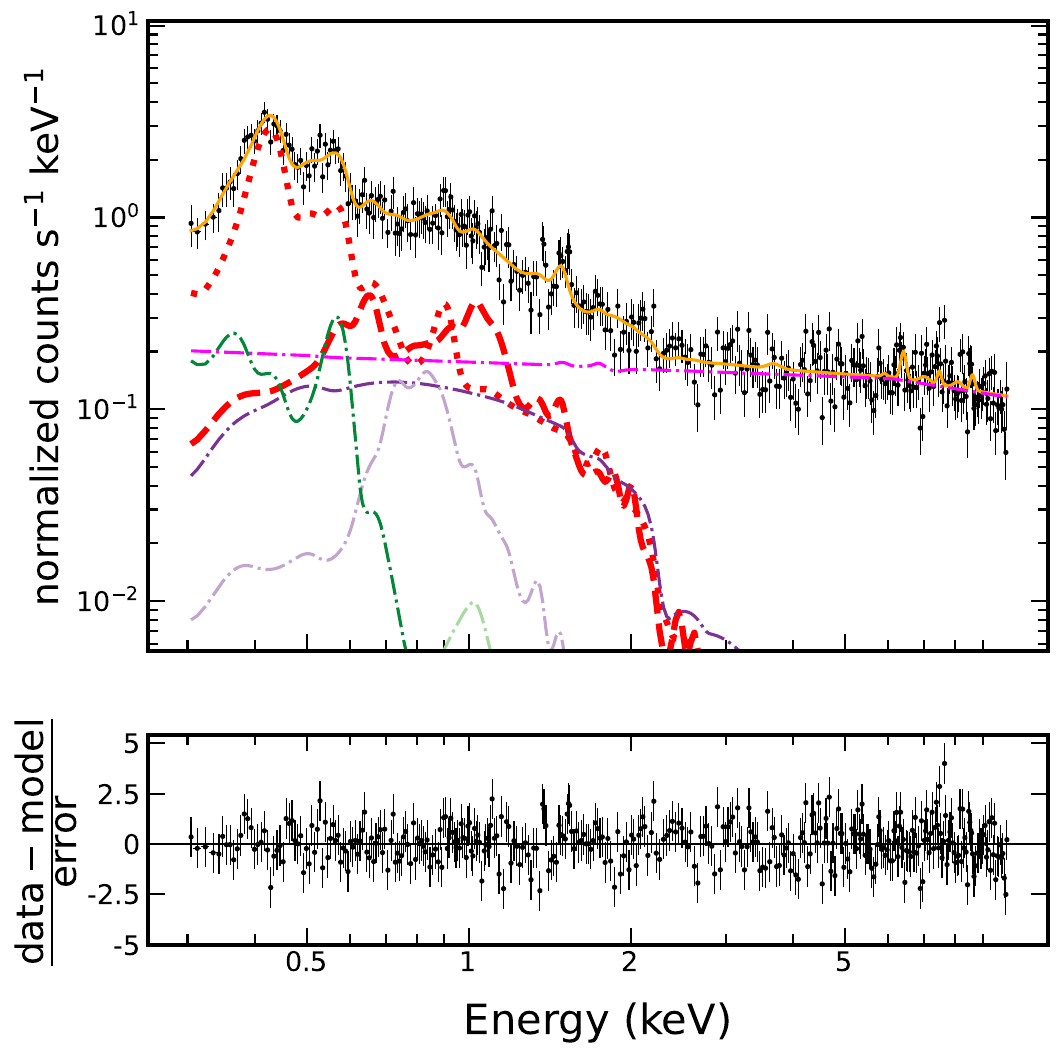}
    \caption{Spectrum of the diffuse emission surrounding WR6. \textit{Top-left}: \texttt{tbabs*(vapec+vapec)}, \textit{Top-right}: \texttt{tbabs*(vpshock+vpshock)} with fixed abundances, \textit{Bottom}: \texttt{tbabs*(vpshock+vpshock)} with the abundances left free to vary, as described in the text. For all the plots, the source model is shown as two red lines (two components). The other dash-dotted lines represent the background, demonstrating that the flat powerlaw is the most significant component of the particle background, while the two thermal dashdot components represent the galactic halo. Finally, the powerlaw shows the unresolved Cosmic X-ray Background. The orange line is the combined model. The models displayed are extracted from the median values of the last 2000 steps of the chain run of \texttt{emceee}. The spectra have also been rebinned for a clearer presentation.
    \label{fig:Spectrum_2vapec_extraN}}
\end{figure*}

 We started modeling the diffuse emission with a single absorbed (TBabs, \cite{Wilms2000}) VAPEC model, obtaining a CSTAT/DOF (where DOF=degree of freedom) close to 1.07. As presented also in \cite{Toala2012}, we tried to add another thermal component. Setting one component with abundances fixed to those adopted by \cite{Chu2003} and \cite{Toala2012}, namely, C/C$_{\odot}$=0.1, N/N$_{\odot}$=1.6, O/O$_{\odot}$=0.13, Mg/Mg$_{\odot}$=0.13, Ne/Ne$_{\odot}$=0.22, and Fe/Fe$_{\odot}$=0.13, we also added another VAPEC with only the N abundance free to vary. This modeling provides a better fit (CSTAT/DOF=1.02) and the spectrum is shown in Figure \ref{fig:Spectrum_2vapec_extraN}. The model is based on two collisional equilibrium thermal plasma components which have respectively as temperature kT$_{1}=1.10_{-0.10}^{+0.11}$ keV and kT$_{2}=0.103_{-0.007}^{+0.005}$ keV. This second component displays a clearly supersolar N abundance (N/N$_{\odot}$=$9.1_{-1.1}^{+0.6}$). The best-fit parameters are shown in Table \ref{tab:BF_2VAPEC}.
 
 \begin{table}[]
\centering
\caption{Best-fit values obtained with a double VAPEC model. The normalization factors ("factor") of the instrumental and sky background are also expressed.\label{tab:BF_2VAPEC}}
\begin{tabular}{cc}
Model&\tiny{constant*TBabs*(vapec + vapec)}\\[1ex]%
\hline%
Parameter&\\[1ex]%
\hline%
Instrumental background&\\[1ex]%
factor&$0.153_{-0.002}^{+0.003}$\\[1ex]%
\hline%
Source&\\[1ex]%
\hline%
N$\_$H (10$^{22}$ cm$^{-2}$)&$0.05_{-0.03}^{+0.05}$\\[1ex]%
kT (keV)&$1.10_{-0.10}^{+0.11}$\\[1ex]%
Normalization&$0.0058_{-0.0012}^{+0.0014}$\\[1ex]%
kT (keV)&$0.103_{-0.007}^{+0.005}$\\[1ex]%
C/C$_{\odot}$&$0.07_{-0.05}^{+0.27}$\\[1ex]%
N/N$_{\odot}$&$9.1_{-1.1}^{+0.6}$\\[1ex]%
Normalization&$0.014_{-0.005}^{+0.012}$\\[1ex]%
\hline%
Sky background&\\[1ex]%
factor&$0.46_{-0.07}^{+0.07}$\\[1ex]%
\hline%
Statistic&893/877=1.02\\[1ex]%
\end{tabular}
\begin{flushleft}
\small
 Normalization is expressed as $10^{-14}\dfrac{\int n_{e}n_{H}dV}{4\pi D^{2}}$, where n$_{e}$ is the electron density of the plasma (cm$^{-3}$), n$_{H}$, is the hydrogen density (cm$^{-3}$) and D (cm) is the distance to the source. 
\end{flushleft}
\end{table}

 The result is quite different if two non-equilibrium collisional shock models are employed (VPSHOCK, \cite{Borkowski2001}) and the same abundance ratios are adopted. Again, we considered the galactic absorption of these components using the TBabs model. Despite the fact that the CSTAT statistics was improved, whilst obtaining the best-fit value so far, the temperatures seem a bit unrealistic, being well above 1 keV for both components (Table \ref{tab:BF_2VPSHOCK_fixed}).

\begin{table}[]
\centering
\caption{Best-fit values obtained with a double VPSHOCK model. The normalization factors ("factor") of instrumental and sky background are also expressed.\label{tab:BF_2VPSHOCK_fixed}}
\begin{tabular}{cc}
Model&\tiny{constant*TBabs*(vpshock + vpshock)}\\[1ex]%
\hline%
Instrumental background&\\[1ex]%
factor&$0.154_{-0.003}^{+0.003}$\\[1ex]%
\hline%
Source&\\[1ex]%
\hline%
N$\_$H (10$^{22}$ cm$^{-2}$)&$0.08_{-0.03}^{+0.03}$\\[1ex]%
kT (keV)&$1.4_{-0.7}^{+1.8}$\\[1ex]%
Tau$_u$ (10$^{10}$ cm$^{-3}$ s)&$0.16_{-0.03}^{+0.04}$\\[1ex]%
Normalization&$0.005_{-0.002}^{+0.004}$\\[1ex]%
kT (keV)&$2.1_{-0.9}^{+3.1}$\\[1ex]%
C/C$_{\odot}$&$0.4_{-0.4}^{+3.2}$\\[1ex]%
N/N$_{\odot}$&$0.3_{-0.2}^{+2.1}$\\[1ex]%
Tau$_u$ (10$^{10}$ cm$^{-3}$ s)&$40_{-30}^{+110}$\\[1ex]%
Normalization&$0.0010_{-0.0005}^{+0.0007}$\\[1ex]%
\hline%
Sky background&\\[1ex]%
factor&$0.40_{-0.06}^{+0.05}$\\[1ex]%
\hline%
Statistic&873/875=1.0\\%
\end{tabular}
\begin{flushleft}
\small
 Normalization is expressed as $10^{-14}\dfrac{\int n_{e}n_{H}dV}{4\pi D^{2}}$, where n$_{e}$ is the electron density of the plasma (cm$^{-3}$), n$_{H}$, is the hydrogen density (cm$^{-3}$) and D (cm) is the distance of the source
\end{flushleft}
\end{table}
  
 We then tested again a double non-equilibrium model with the abundances of N, O, Ne, Mg, and Fe free to vary in one component. We have chosen to let vary these specific ions since these are the lines that are more prominent in our spectra. In the second model component, we left only C and N free to vary. Our aim here was testing the assumption made by \cite{Chu2003} and \cite{Toala2012}, using  two non-equilibrium models instead. In the first instance, we obtained a fit with the two components inverted, due to the overly high degeneracy among the parameters. To solve this issue, we started from the best fit obtained with the fixed abundance and we launched another fit run, allowing the abundances listed above to be free to vary.

 We retrieved abundances similar to those originally adopted by \cite{Chu2003}. The final picture is a two-phase gas: one presenting lower temperature and very low ionization timescale, while the other presents a high temperature and a slightly higher ionization timescale. The plasma is far from reaching equilibrium in both components, as per the definition of ionization timescale given in \cite{Xspec,Borkowski1994,Borkowski2001}. From Figure \ref{fig:Spectrum_2vapec_extraN}, we observe that the hot and high ionization timescale component is much weaker. In performing the same exercise of letting free the abundances in the double VAPEC model, we obtained CSTAT/DOF=1.02. Nevertheless, with the double VAPEC we recover the temperatures of \cite{Toala2012}, but obtaining a much higher N abundance (N/N$_{\odot}=9.3_{-0.9}^{+0.5}$). Given the difference of CSTAT/DOF is around 1\% between the two models, we suggest that deeper data with much higher spectral resolution would be needed to finally determine whether the difference between the two models is real or the effect is mainly related to the lack of spectral resolution.  We also ran again the fits starting from 0.2 keV instead of 0.3 keV, but the results are substantially the same. We opted to continue using the results from the spectra starting from 0.3 keV, given the calibration of eROSITA is more stable and known above 0.3 keV.

\begin{table}[]
\centering
\caption{Best-fit values obtained with a double VPSHOCK model, with N, O, Ne, Mg and Fe free to vary in the first component and only C and N are free to vary in the second. The normalization factors ('factor') of instrumental and sky background are also expressed.\label{tab:BF_2VPSHOCK_free}}
\begin{tabular}{cc}
Model&\tiny{constant*TBabs*(vpshock + vpshock)}\\[1ex]%
\hline%
Instrumental background&\\[1ex]%
factor&$0.154_{-0.003}^{+0.003}$\\[1ex]%
\hline%
Source&\\[1ex]%
\hline
N$\_$H (10$^{22}$ cm$^{-2}$)&$0.14_{-0.05}^{+0.05}$\\[1ex]%
kT (keV)&$0.8_{-0.3}^{+0.8}$\\[1ex]%
C/C$_{\odot}$&$0.3_{-0.3}^{+1.0}$\\[1ex]%
N/N$_{\odot}$&$3_{-2}^{+3}$\\[1ex]%
O/O$_{\odot}$&$0.3_{-0.2}^{+0.2}$\\[1ex]%
Ne/Ne$_{\odot}$&$0.5_{-0.4}^{+0.8}$\\[1ex]%
Mg/Mg$_{\odot}$&$0.05_{-0.03}^{+0.14}$\\[1ex]%
Fe/Fe$_{\odot}$&$0.2_{-0.2}^{+1.3}$\\[1ex]%
Tau$_u$ (10$^{10}$ cm$^{-3}$ s)&$0.11_{-0.03}^{+0.03}$\\[1ex]%
Normalization&$0.008_{-0.004}^{+0.008}$\\[1ex]%
kT&$2_{-1}^{+3}$\\[1ex]%
C/C$_{\odot}$&$1_{-1}^{+4}$\\[1ex]%
N/N$_{\odot}$&$0.04_{-0.02}^{+0.09}$\\[1ex]%
Tau$_u$ (10$^{10}$ cm$^{-3}$ s)&$60_{-40}^{+290}$\\[1ex]%
Normalization&$0.0013_{-0.0006}^{+0.0012}$\\[1ex]%
\hline%
Sky background&\\[1ex]%
factor&$0.40_{-0.11}^{+0.08}$\\[1ex]%
\hline%
Statistic&874/869=1.01\\[1ex]%
\end{tabular}
\begin{flushleft}
\small
 Normalization is expressed as $10^{-14}\dfrac{\int n_{e}n_{H}dV}{4\pi D^{2}}$ where n$_{e}$ is the electron density of the plasma (cm$^{-3}$), n$_{H}$ is the hydrogen density (cm$^{-3}$) and D (cm) is the distance of the source
\end{flushleft}
\end{table}
 
 In this context, given the limited spectral resolution of  present-day X-ray instruments (especially on extended sources), \cite{Kavanagh2020} discusses how both APEC and non-equilibrium models would not serve as the most suitable models to describe the complex environments around bubbles and superbubbles. Since, with both models, we have a CSTAT/DOF value that is very close to 1 and given the very small difference of statistics between the two models tested, we conclude that the argument of \cite{Kavanagh2020} should be valid also around wind-blown bubbles, as in the case of S 308.

\subsection{Galactic search for other WR X-ray shining bubbles}
Given the result obtained for WR6, we looked also at the WR nebulae listed in Table 3 of \cite{Toala2015} and we calculated the flux per each of them obtained with the eRASS:4 dataset. We proceeded as follows. 

We started by downloading the \textit{WISE} image from the IRSA\footnote{\url{https://irsa.ipac.caltech.edu/Missions/wise.html}}  archive for each one of the WR nebulae. Next, for each \textit{WISE} image, we defined a circular region in DS9 matching the boundaries of the nebula. The region size depends on the dimension and the shape of the bubble, which can be irregular and not always clearly visible. In this case, we tried to match the brightest boundary of the bubble visible in IR, keeping the circle centered on the WR star. Then, we masked the point sources inside the extraction region and we extracted a spectrum for each one of these regions. We used a region of the same size to extract the background from a nearby region. In this way, we extracted an independent background spectrum for each bubble. Assuming the double VPSHOCK model with fixed abundance employed on S308  (described in Section \ref{sec:SpectralAnalysisS308}), we measured the X-ray fluxes at the position of the optical/IR bubbles listed in Table 3 of \cite{Toala2015}. The results are shown in Table \ref{tab:Bubbles_Fluxes} and the extraction regions in Figures \ref{fig:WR_7_38} and  \ref{fig:WR_40_101}.

\begin{table}[]
\centering
\caption{3$\sigma$ X-ray fluxes measured with eROSITA for the bubbles listed in Table 3 of \cite{Toala2015}. The bubbles are located between 180 and 360 degrees of galactic longitudes. The first column displays the identifier of the WR star and the second the flux from the source region. \label{tab:Bubbles_Fluxes}}
\begin{tabular}{cc}
WR identifier & Flux ($10^{-13}$ erg s$^{-1}$ cm$^{-2}$)  \\[1ex]%

\hline%
WR7&$<3.0$\\[1ex]%
WR8&$<1.9$\\[1ex]%
WR16&$<5.4$\\[1ex]
WR18&$<2.6$\\[1ex]%
WR22&$<42$\\[1ex]%
WR23&$<19$\\[1ex]%
WR30&$<4.7$\\[1ex]%
WR31a&$<0.75$\\[1ex]
WR35&$<2.1$\\[1ex]%
WR35b&$<14$\\[1ex]%
WR38&$<5$\\[1ex]%
WR40&$<7$\\[1ex]%
WR52&$<7$\\[1ex]%
WR54&$<20$\\[1ex]%
WR55&$<4.6$\\[1ex]%
WR68&$<7$\\[1ex]%
WR75&$<76$\\[1ex]%
WR85&$<12$\\[1ex]%
WR86&$<14$\\[1ex]%
WR94&$<16$\\[1ex]%
WR95&$<42$\\[1ex]%
WR101&$<19$\\[1ex]%
\hline
\end{tabular}
\end{table}
For simplicity and given the minimal differences among Table \ref{tab:BF_2VPSHOCK_fixed} and Table \ref{tab:BF_2VPSHOCK_free}, we proceeded with the double VPSHOCK model with fixed abundance to extract the flux of the 22 bubbles. In addition, thanks to explicit background modeling, we took into account the background in the flux estimation. Therefore, the fluxes reported in Table \ref{tab:Bubbles_Fluxes} at 3$\sigma$ confidence level are only from the source: this allows us to put upper limits on the source flux for each  bubble. The values at 3$\sigma$ confidence level are obtained taking 99.8\% of the posterior probability distribution of the last 2000 steps of the chains. We computed the significance checking whether S/N>3, where S is the counts in the source region while and N is total count in the background region. Both S and N values are normalized for the region area.  In cases where the ratio is higher than 3, we claimed a 3 sigma detection. Otherwise, we estimated the 3$\sigma$ flux upper limit. We do not detect any of the 22 bubbles above the background level.

\section{Discussion \label{sec:Discussion}}
\subsection{Morphological and spectral analysis of S308}
 Thanks to the all-sky dataset provided by eROSITA, we present the first complete image of the wind-blown bubble S308 in X-rays, together with the first spectral analysis of the object in its entire extent. In addition, we were able to independently select the background region, contrary to what has been possible so far with only \textit{XMM-Newton} pointed observations. From our analysis, the emission of the nebula can be described by a two component non-equilibrium thermal model (kT$_{1}=0.8_{-0.3}^{+0.8}$ keV and kT$_{2}=2_{-1}^{+3}$ keV), presenting very low column density N$\_$H$=0.14_{-0.05}^{+0.05}$ 10$^{22}$ cm$^{-2}$ and an enrichment of N in the first component. The abundance values are remarkably in accordance with those found by \cite{Chu2003}. However, our temperatures do not agree with those of  \cite{Toala2012} when  the double VPSHOCK model is employed, but they agree if the double VAPEC shown in Table \ref{tab:BF_2VAPEC} is used instead. Given the rather unrealistic abundance of N in the second component and the marginal improvement in the statistics given by the double VPSHOCK, we conclude this description is a valid alternative to the double VAPEC first presented by \cite{Toala2012}. We would like to stress that our results are not biased by the background region choice, as opposed to those of \cite{Toala2012}; this has allowed us to leave the column density and the nitrogen abundance  free to vary. Looking at the cold, very soft X-ray peaked emission displayed in Figure \ref{fig:S308_WR6}, to explain the high temperature we found using a non-equilibrium plasma description, we notice how the ionization timescales are small, suggesting the gas is far from the equilibrium \citep{Borkowski1994,Xspec,Borkowski2001}. One possible explanation for such high temperatures and non-equilibrium in both components could be that the shell is expanding in a very low-density medium. This would explain the very low emissivity of the hot component shown in the spectra of Figure \ref{fig:Spectrum_2vapec_extraN}. However, we do agree with \cite{Kavanagh2020} in that most of these models are probably not best suited to describe the complex plasma environment around bubbles, suggesting that new and different models should be employed. For instance, the VAPEC results shown in Table \ref{tab:BF_2VAPEC} have temperature values similar to those of \cite{Toala2012}, but the very high enrichment of N seems a bit unrealistic, as described in Section \ref{sec:SpectralAnalysisS308}. In general, models such as VPSHOCK and VAPEC are more or less sufficient to describe most of the extended sources with the spectral and spatial resolution available with the actual X-ray instruments, but with the advent of forthcoming observatories such as \textit{Arcus} \citep{Arcus2022}, \textit{XRISM} \citep{XRISM2020}, and \textit{Athena} \citep{Athena2013} there will be a need for more refined models. For instance, none of the models considered above take into account photoionization, making the description of this scientific case unlikely to be very accurate given the presence of a very intense illuminating source such as WR6. An intense radiation field provides a viable  explanation for finding a plasma more out of equilibrium in the cold inner component than in the hot one, which we would expect to be farther out with respect to our interpretation.

As visible in Figure \ref{fig:S308_WR6}, and  as previously described in \cite{Chu2003} and \cite{Toala2012}, the structure of the nebula is characterized by a limb of brightened emission, in accordance with our suggestion of hot material expanding in a very low density circumstellar medium. This structure is probably shaped by the intense radiation field of WR6 and powerful winds which push the material further from the star. Looking at the composition of the gas, from our fits, we find a clear signature of N, given a supersolar abundance value in one of the two plasma components, in accordance with \cite{Toala2012}. If the double VAPEC model is adopted, the value is highly supersolar. The presence of a considerable amount of N in the spectrum of S 308 is a clear signature of the enrichment of the outer layers of WR6 that are being progressively expelled during the Wolf-Rayet phase of the star.

\subsection{Search for new X-ray emitting bubbles}
Moreover, we decided to exploit the unique dataset offered by eROSITA to scan the sky in the galactic longitude range of 180 - 360 degrees, with an aim to look for other bubbles similar to S308. We recall that so far X-ray emission has been claimed to be observed only from three other wind-blown bubbles beside S308. Except for NGC 6888 \citep[][and references therein]{Toala2016} around WR136, the other two wind-blown bubbles that are known are located within our search area, specifically: NGC 2359 around the Wolf-Rayet star WR 7 \citep{Zhekov2014} and NGC 3199 around WR 18 \citep{Toala2017}. In this work, we did not detect any of the 22 bubbles at 3$\sigma$ confidence level. However, adopting the double VPSHOCK as the best-fit model, we are able to provide the flux upper limits for each bubble.

\begin{figure}
    \centering
    \includegraphics[scale=0.5]{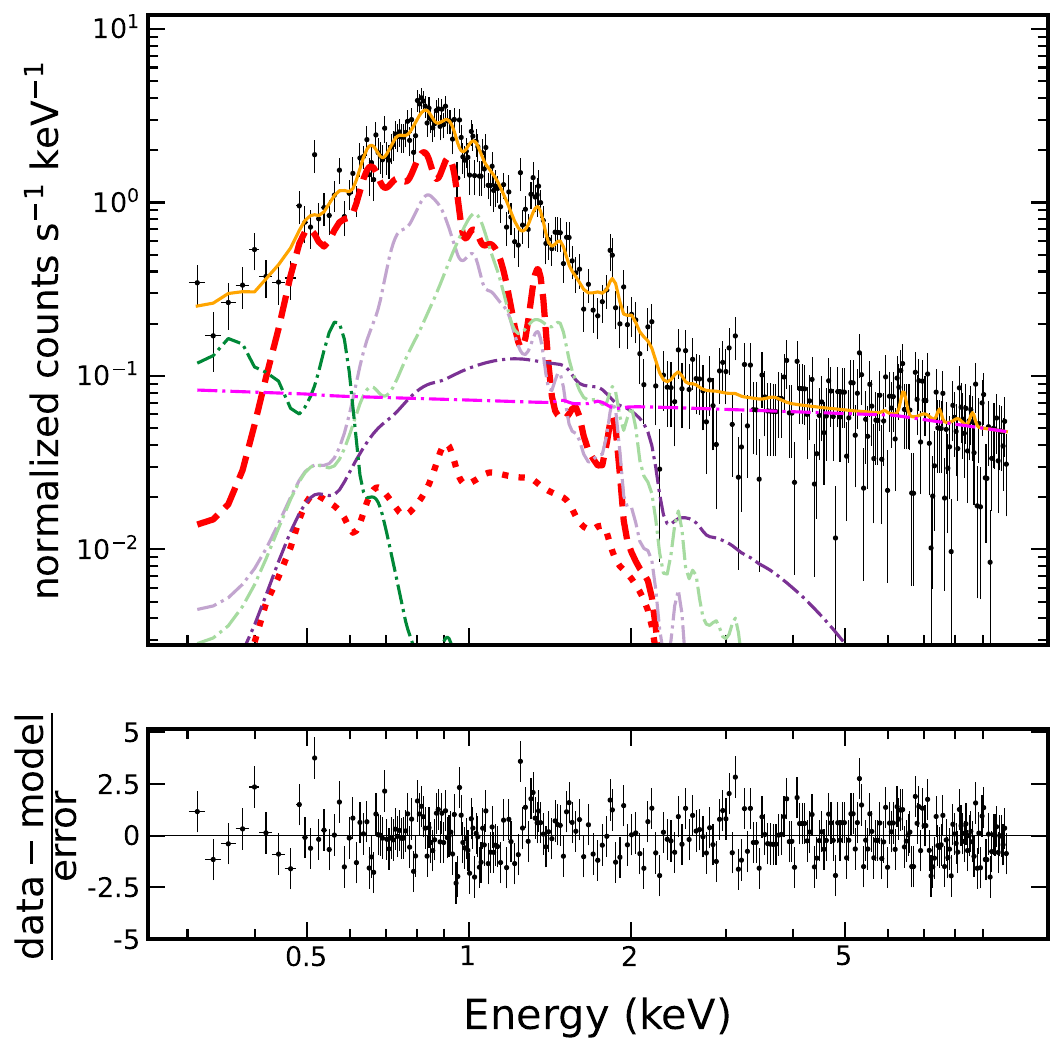}
    \caption{Spectrum of the potential X-ray emitting bubble surrounding WR75. The spectra have been obtained and the lines are labeled as in Figure \ref{fig:Spectrum_2vapec_extraN}.
    \label{fig:Spectrum_WR75}}
\end{figure}

\begin{figure*}
    \centering
    \includegraphics[scale=0.17]{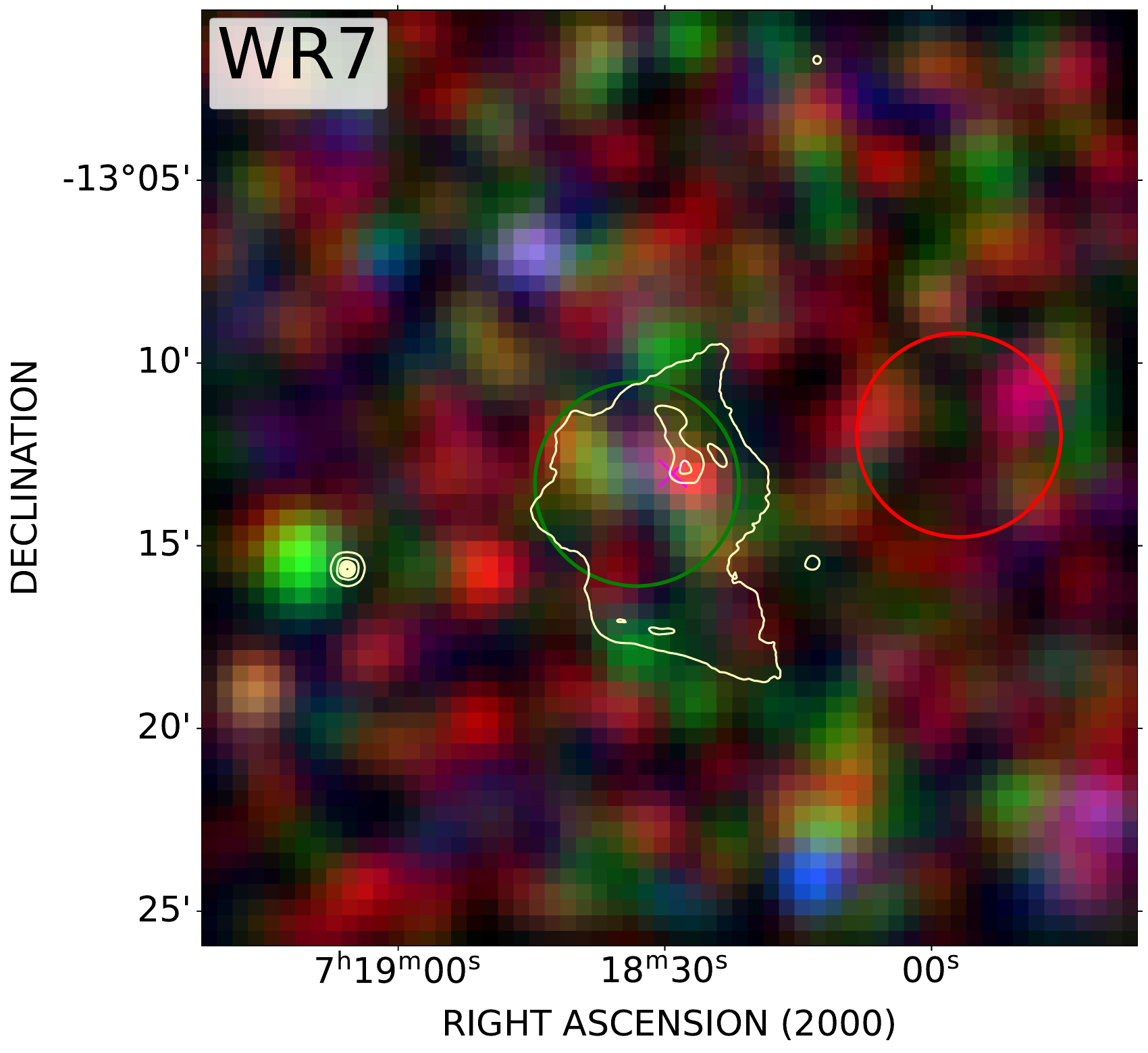}
    \includegraphics[scale=0.17]{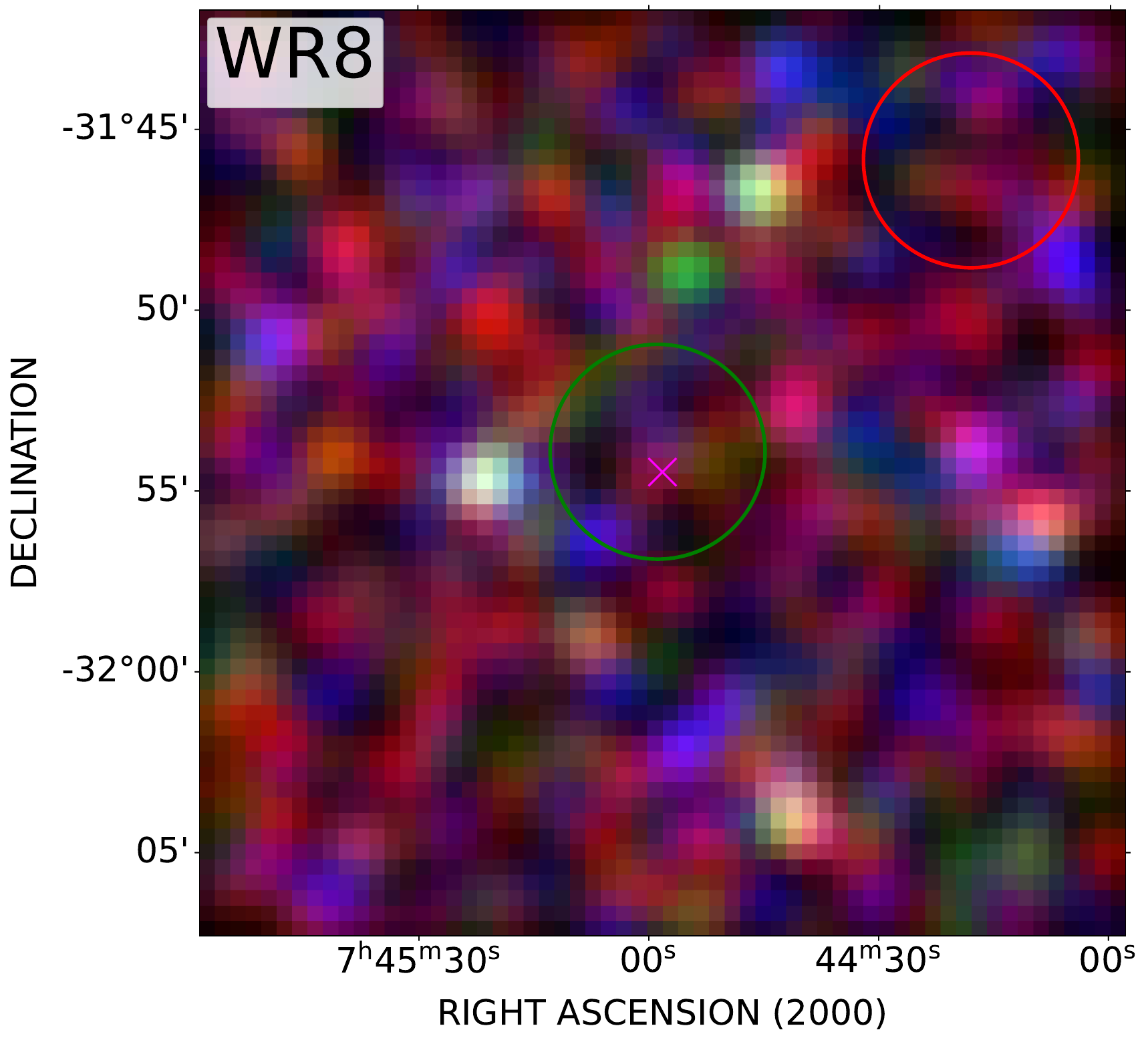}
    \includegraphics[scale=0.17]{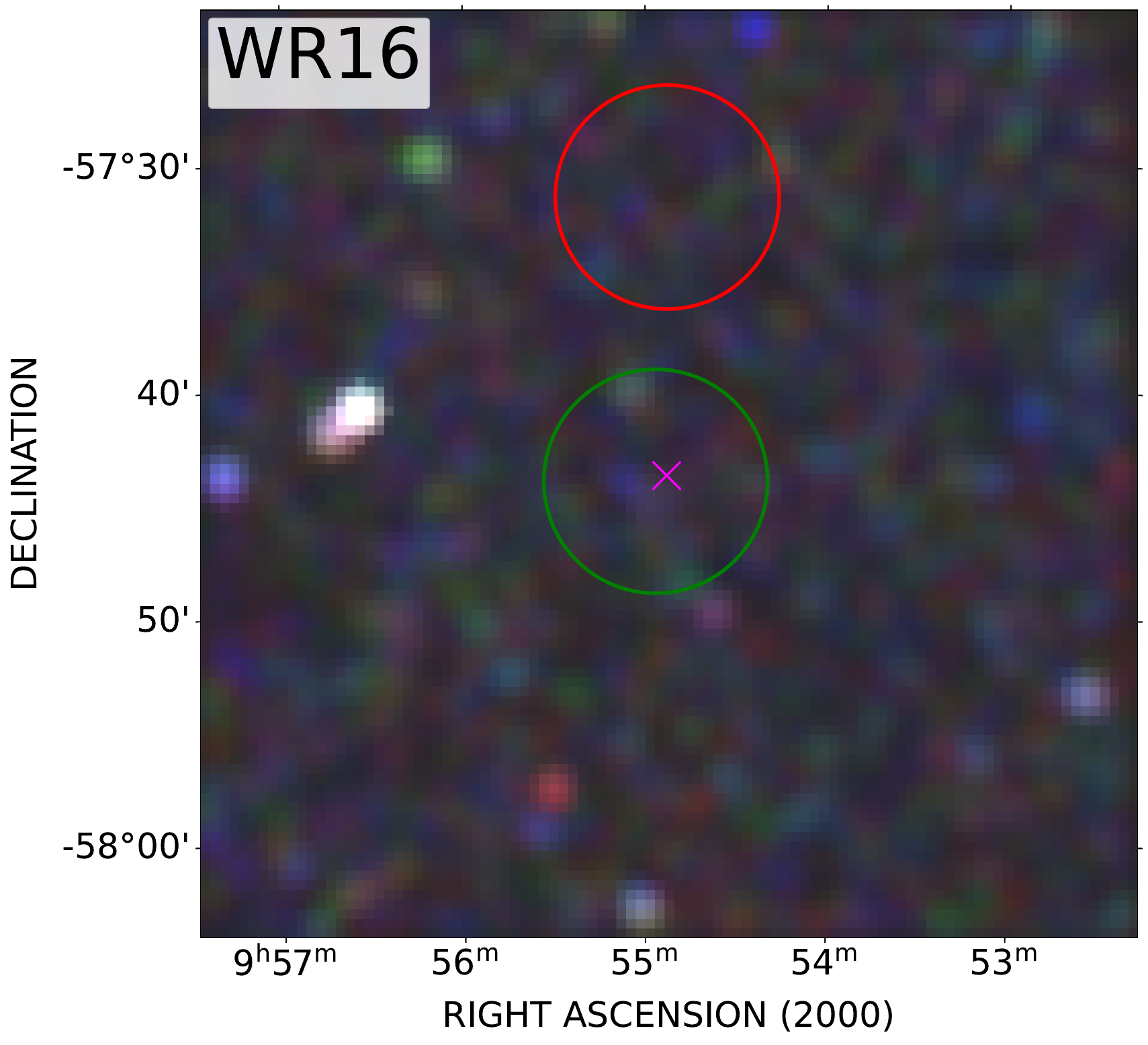}
    \includegraphics[scale=0.17]{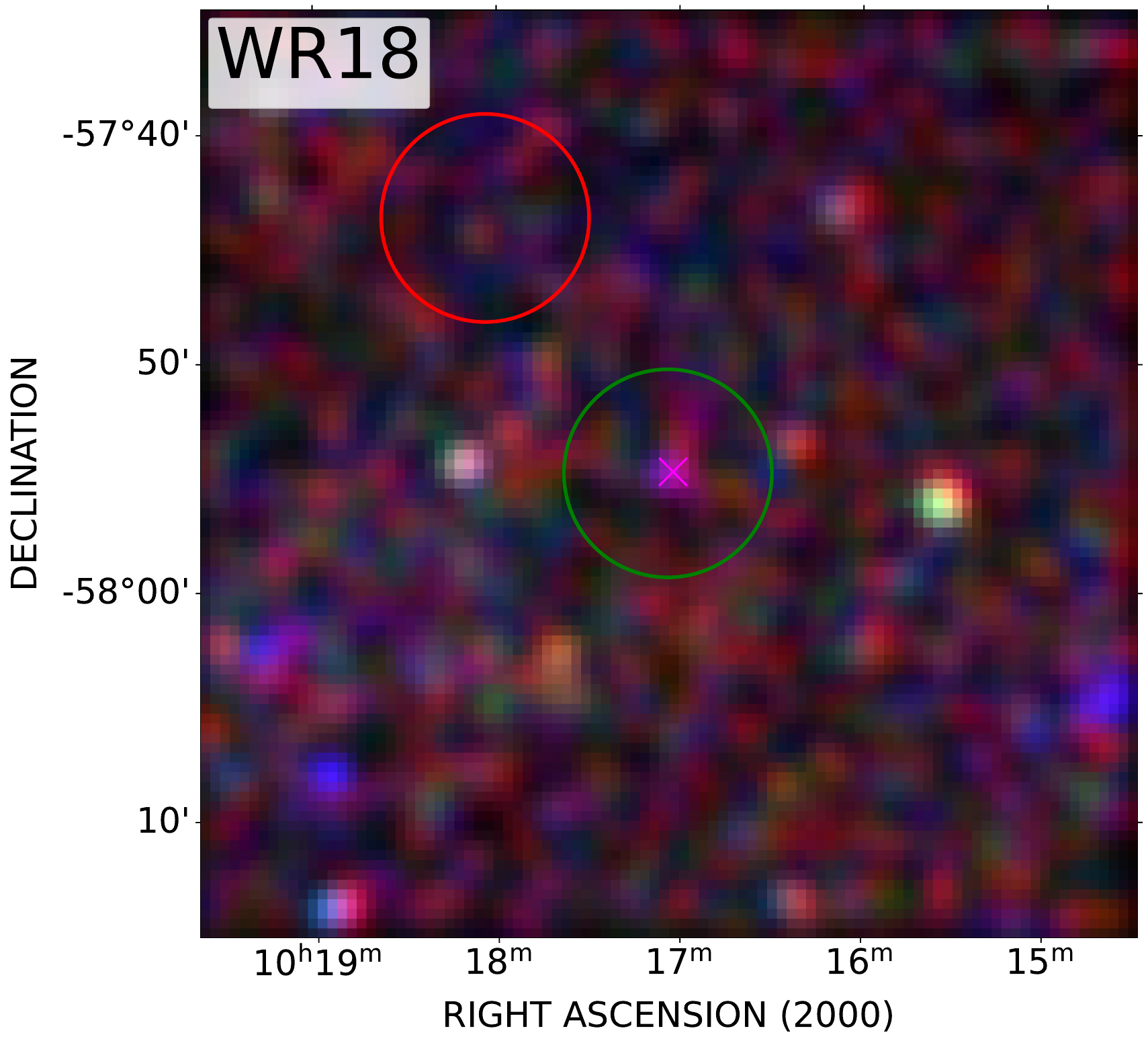}
    \includegraphics[scale=0.17]{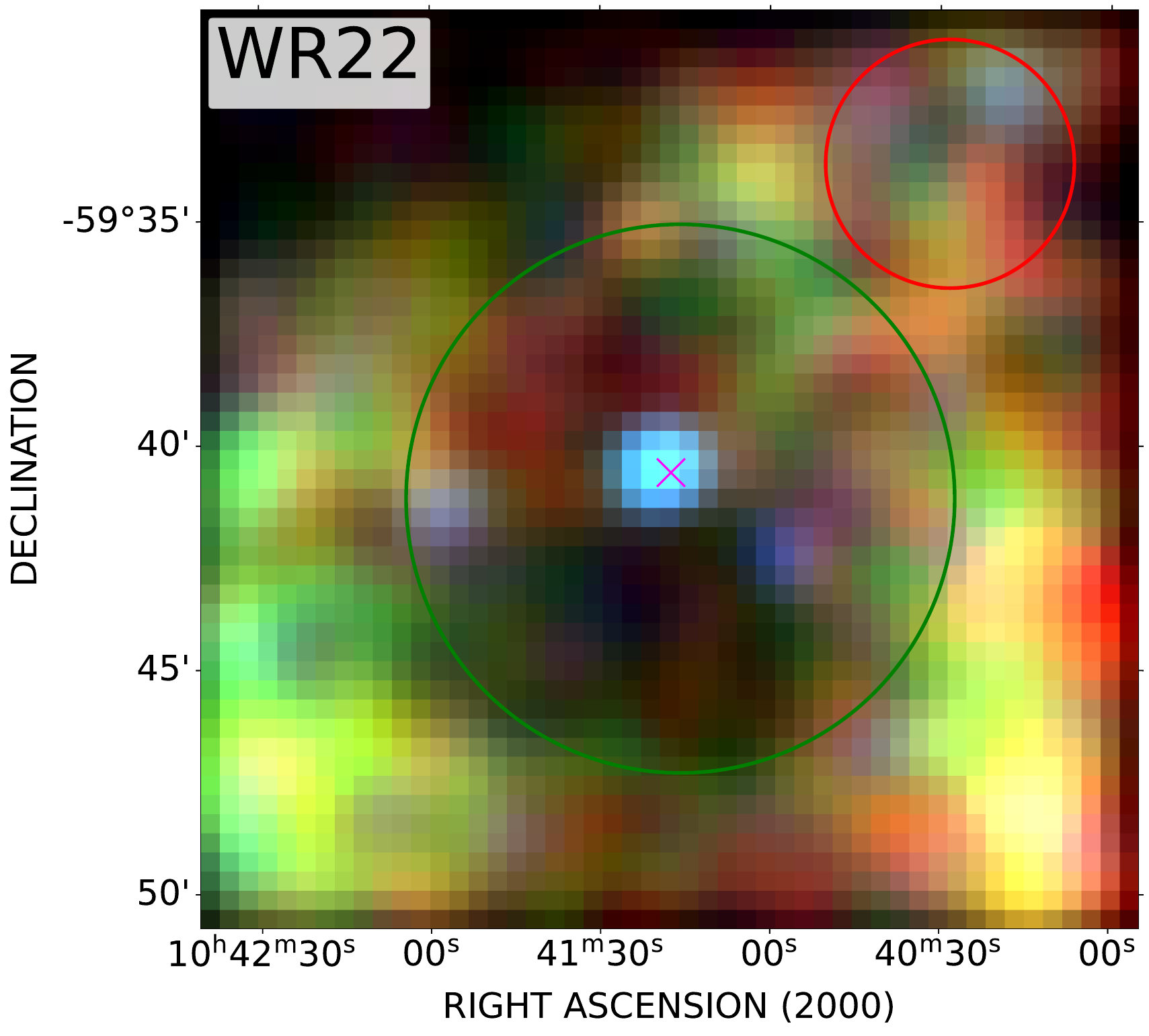}
    \includegraphics[scale=0.17]{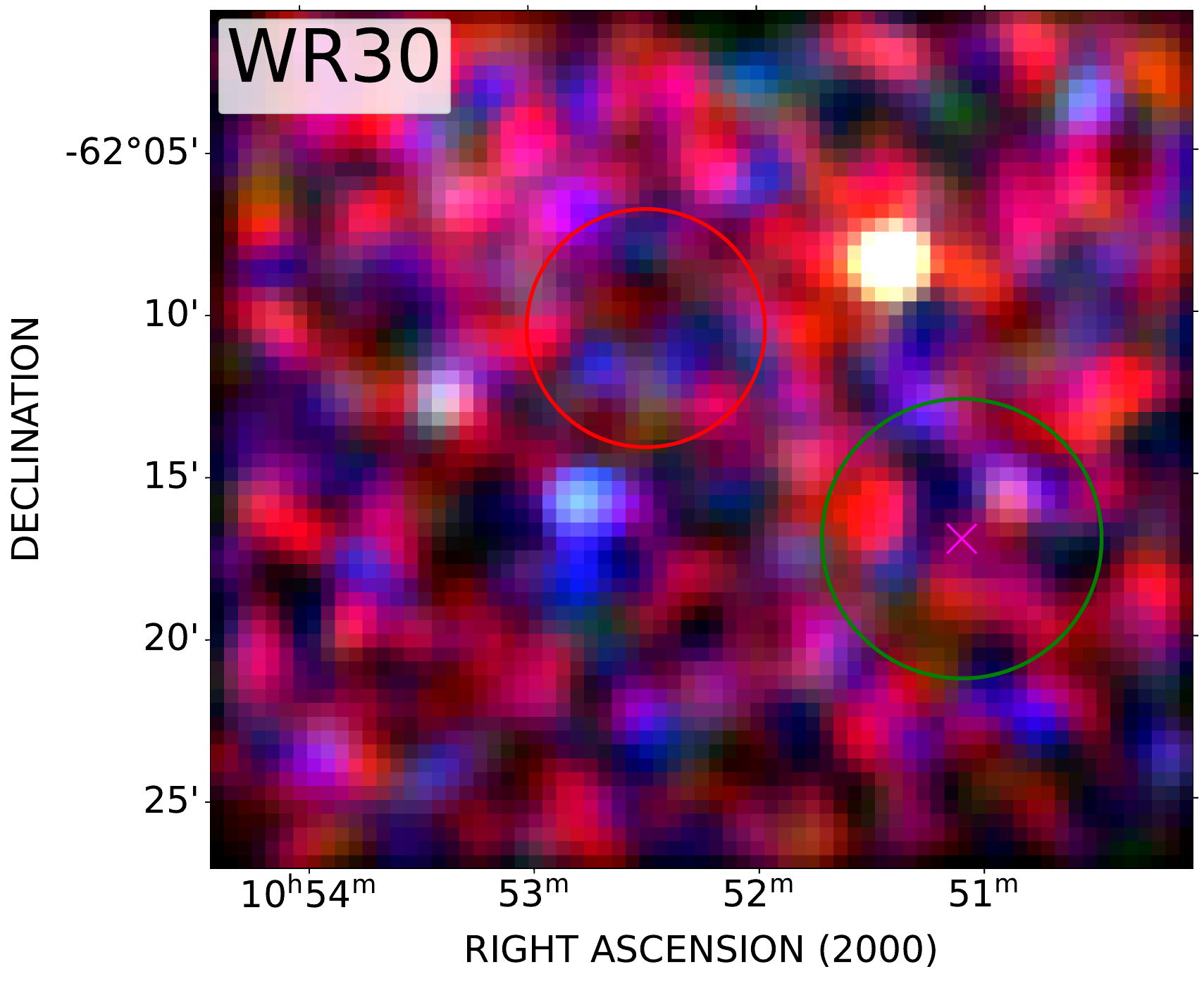}
    \includegraphics[scale=0.17]{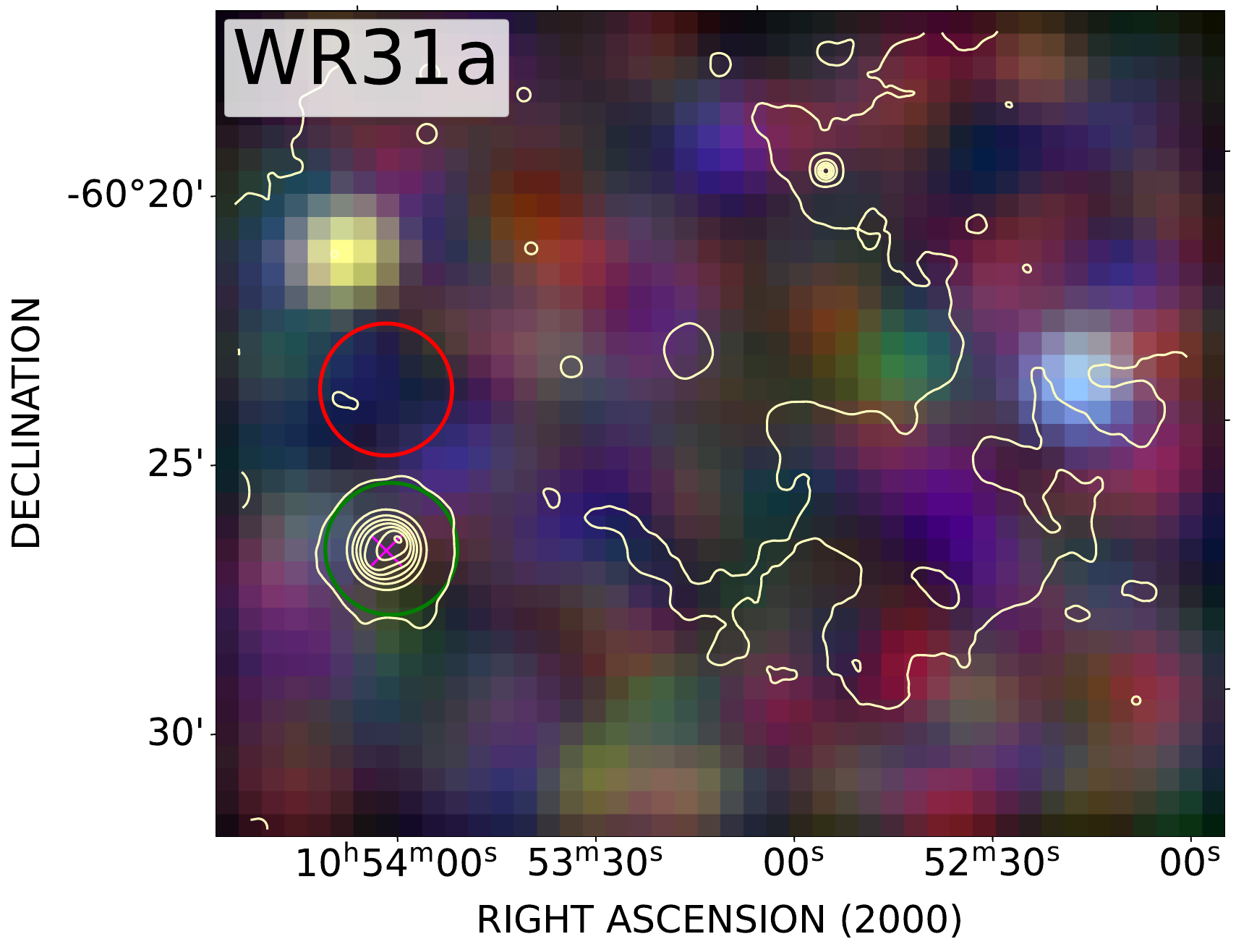}
    \includegraphics[scale=0.17]{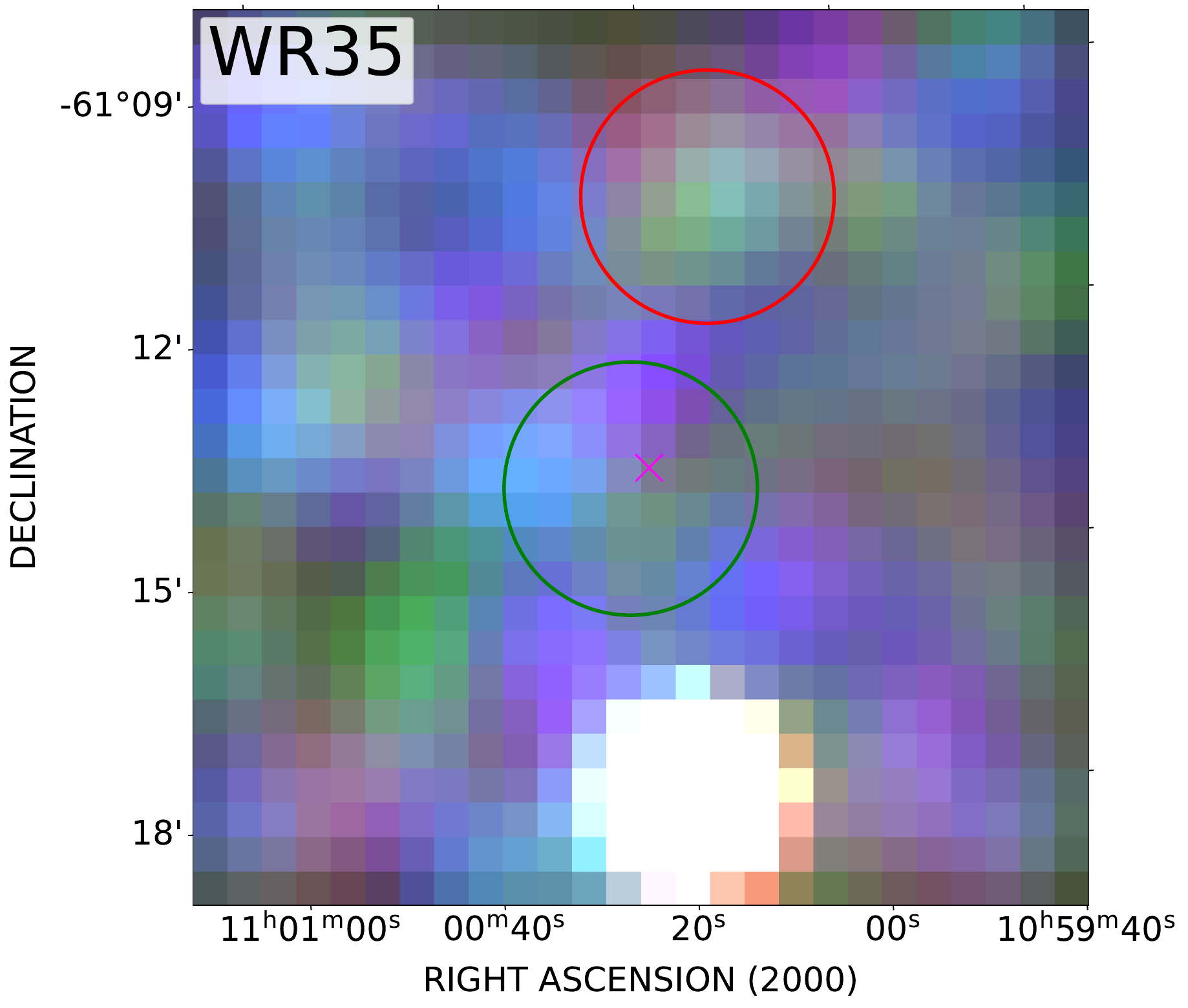}
    \includegraphics[scale=0.17]{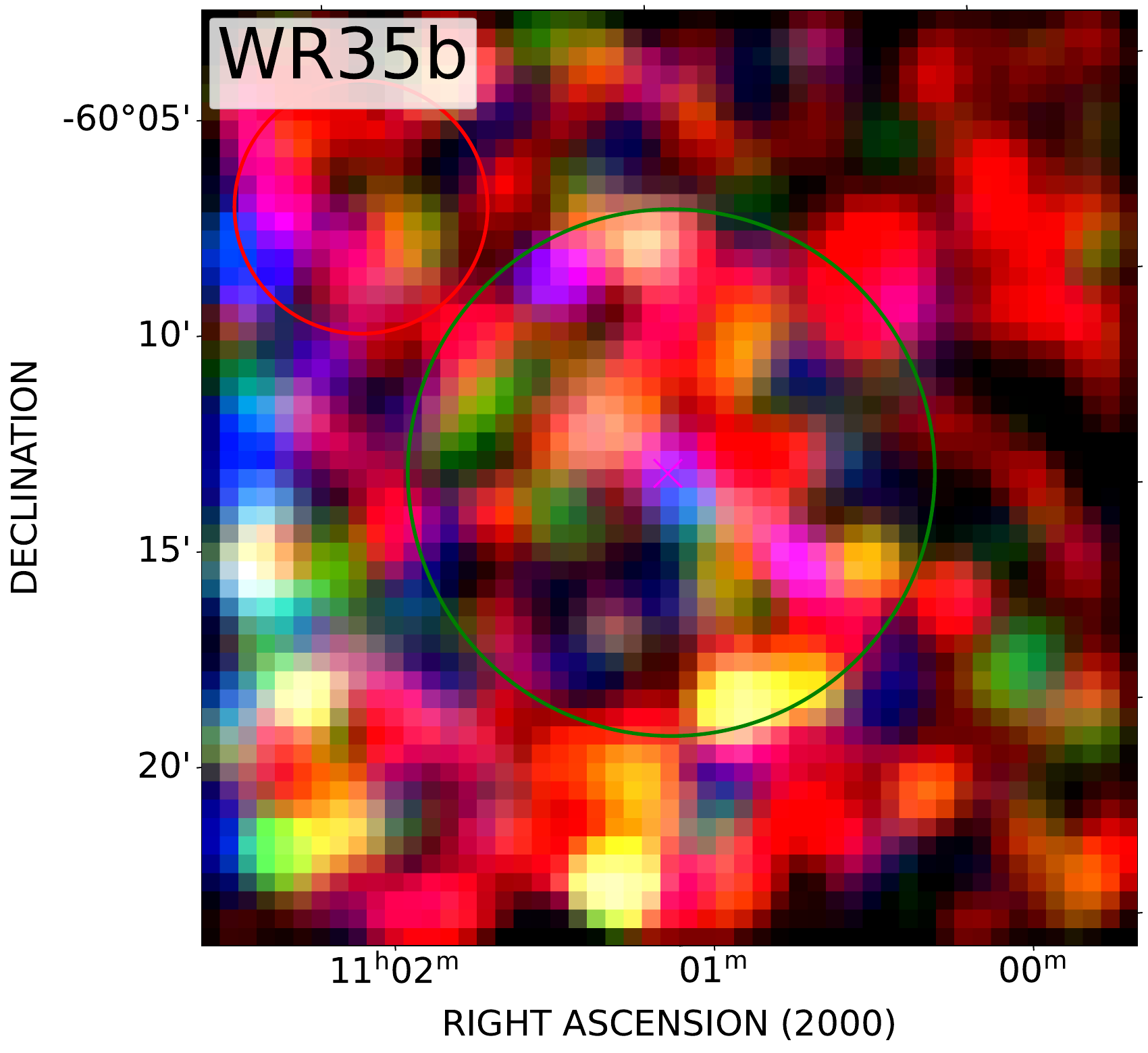}
    \includegraphics[scale=0.17]{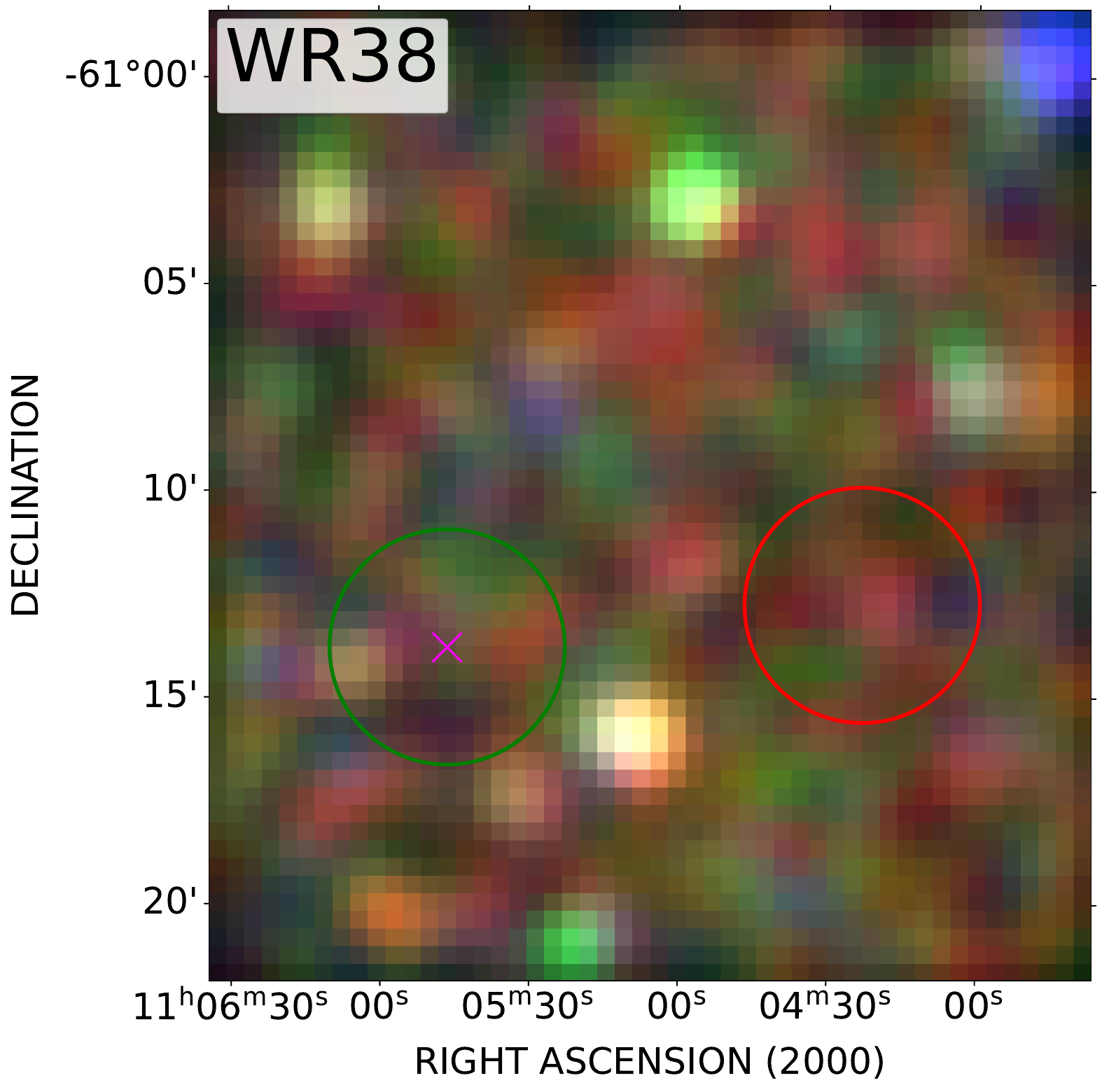}
    \caption{Exposure corrected images in X-rays (Red: 0.2-0.7 keV, Green: 0.7-1.1 keV. Blue: 1.1-10 keV) of the regions surrounding the WR stars listed in Table \ref{tab:Bubbles_Fluxes}. \textit{Top-left}: Corresponding image from WR7 to WR38. We excluded the images of WR23 whose emission is contaminated by surrounding unrelated diffuse emission. In green and red are shown the source and background extraction regions, respectively, employed for the flux estimation (Table \ref{tab:Bubbles_Fluxes}). We show in yellow the WISE contours, only for the most representative bubbles. A magenta cross indicates the position of the WR star.
    \label{fig:WR_7_38}}
\end{figure*}

\begin{figure*}
    \centering
   \includegraphics[scale=0.17]{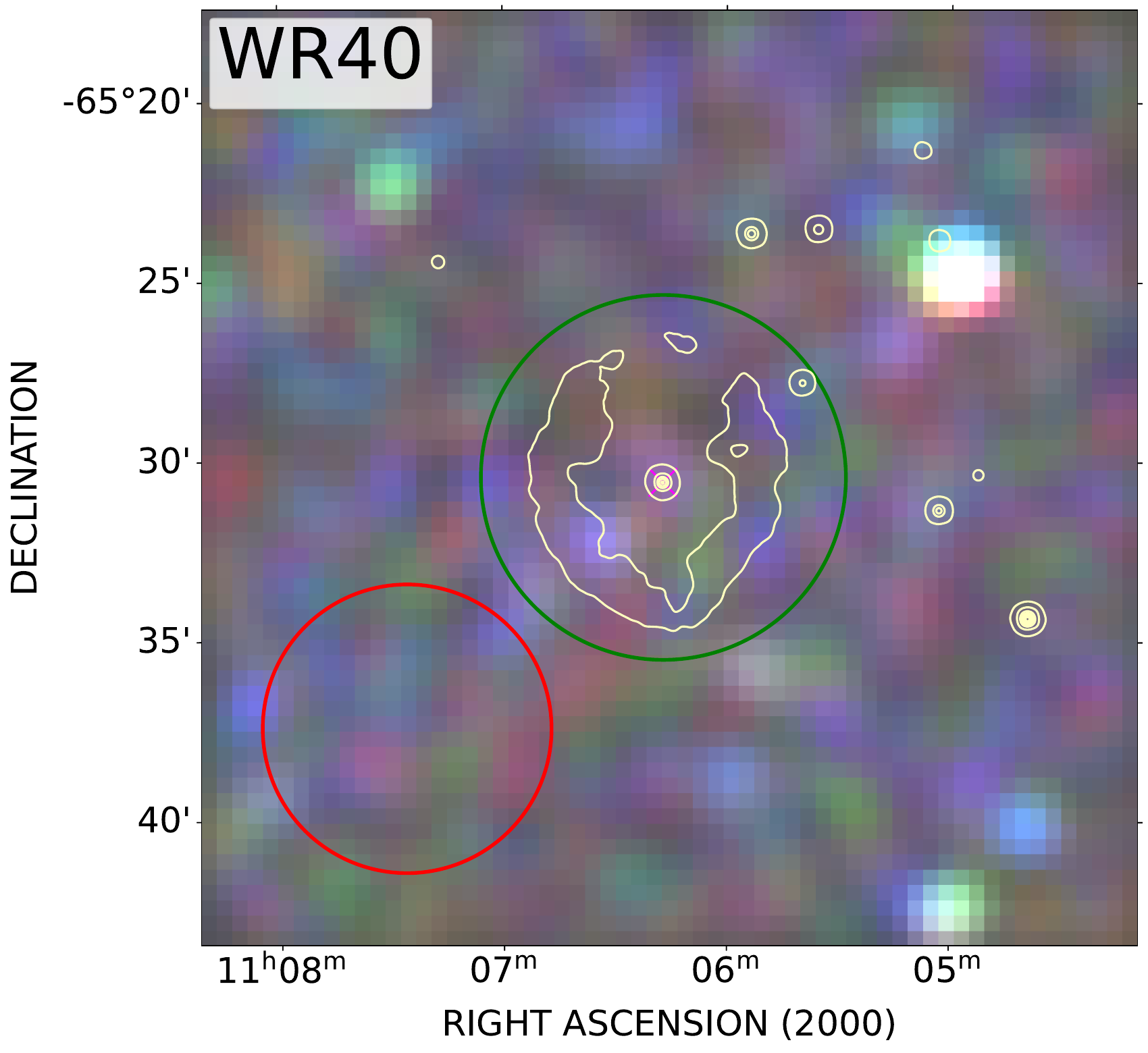}
    \includegraphics[scale=0.17]{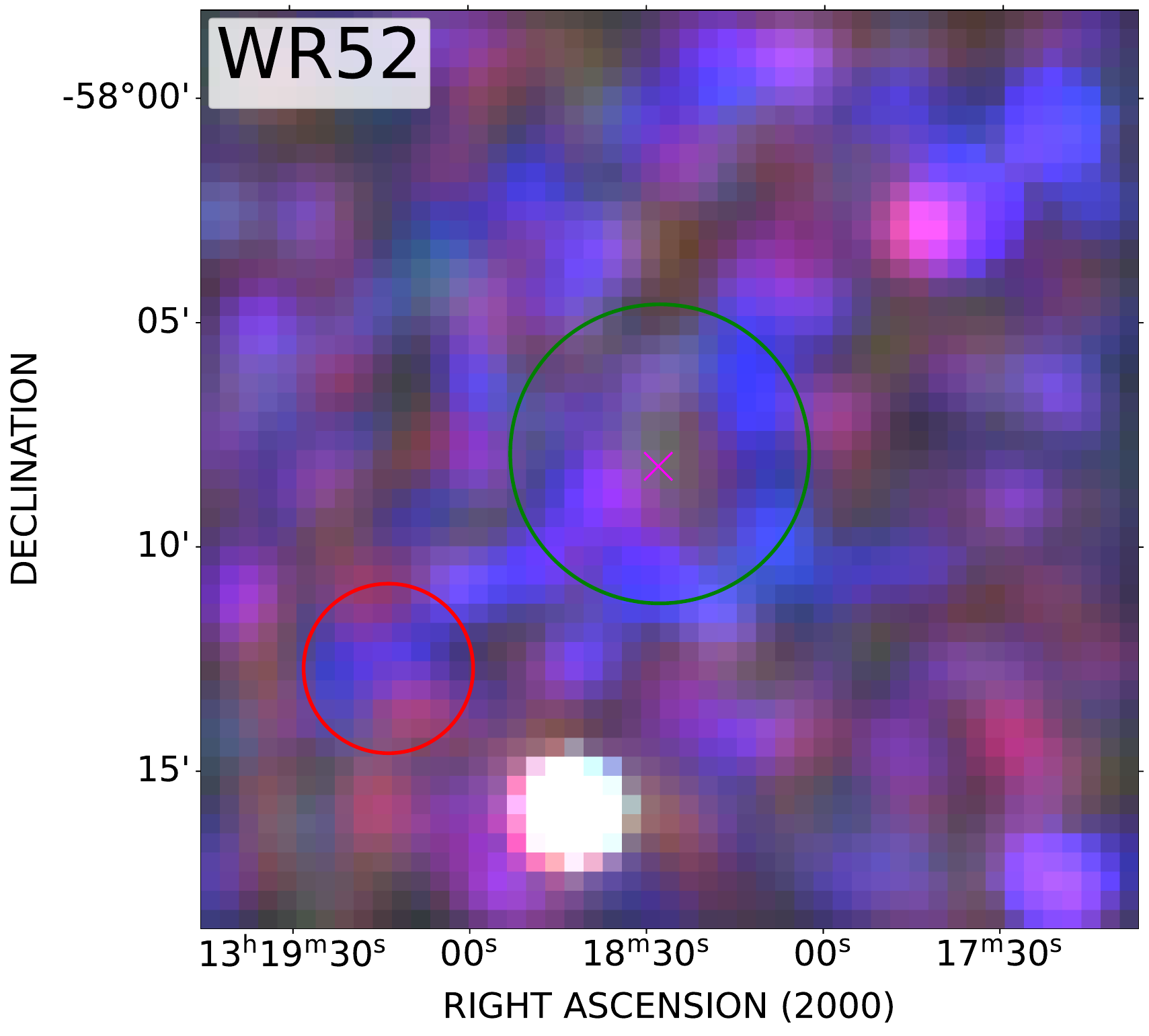}
   \includegraphics[scale=0.17]{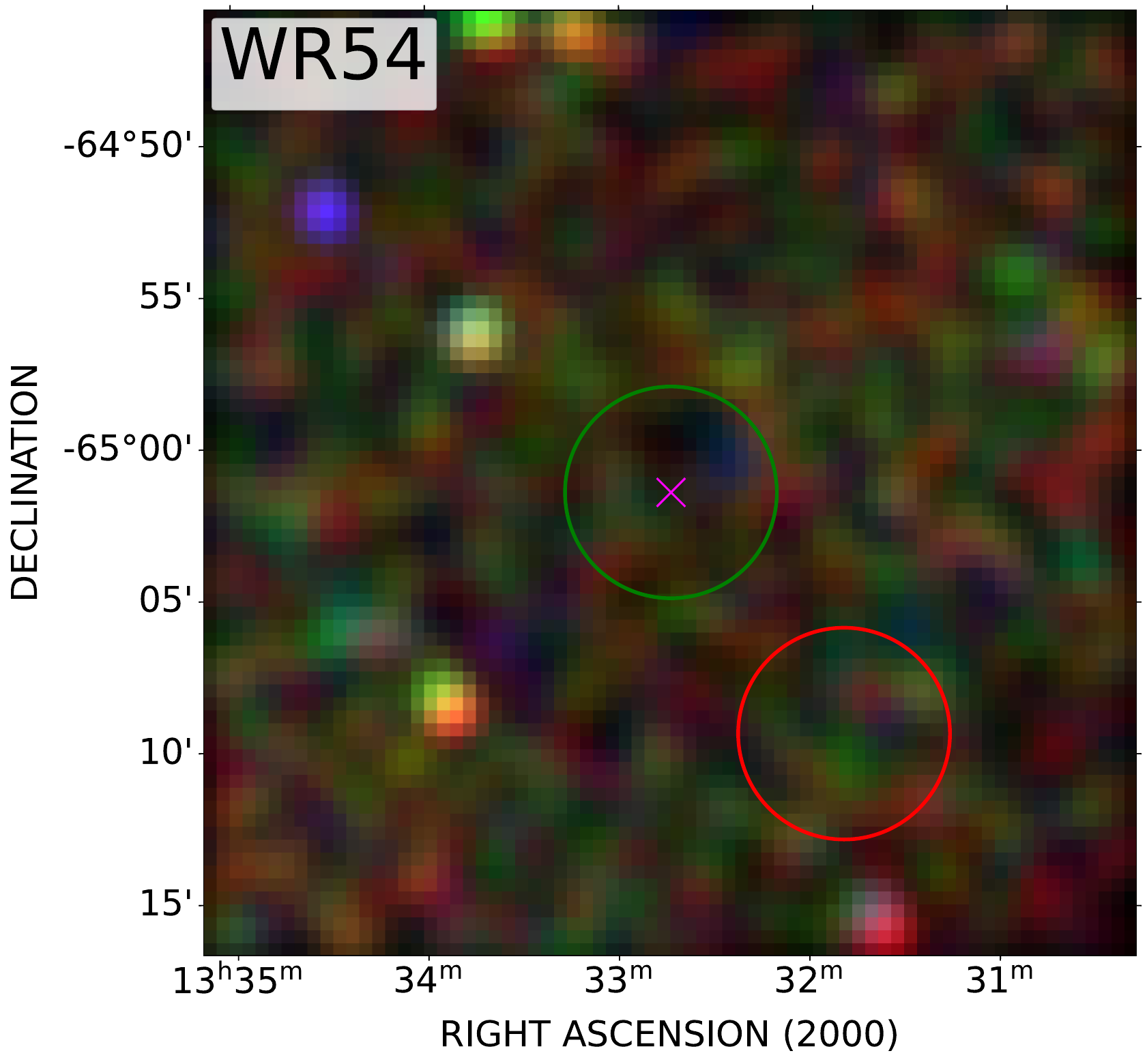}
    \includegraphics[scale=0.17]{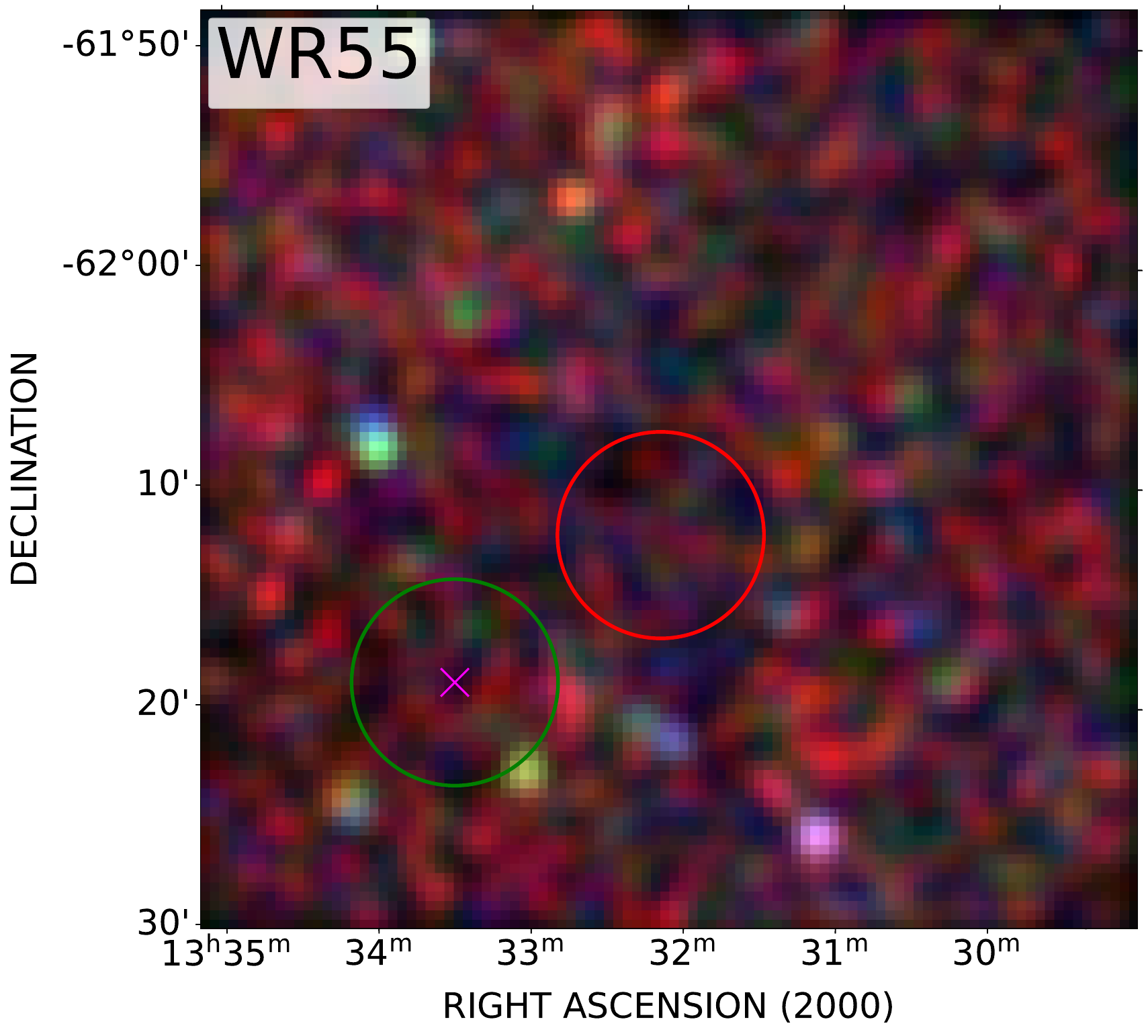}
    \includegraphics[scale=0.17]{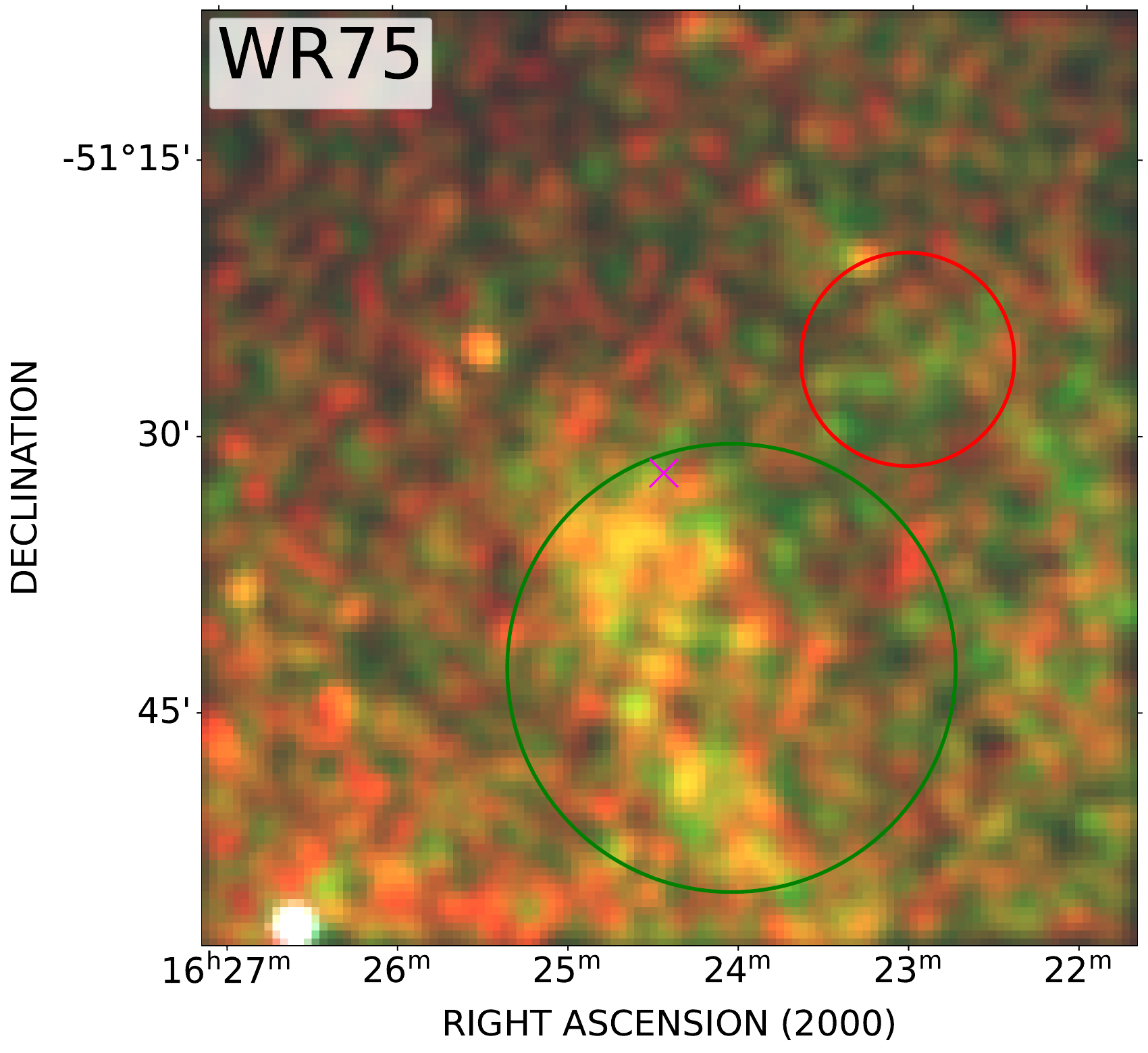}
    \includegraphics[scale=0.17]{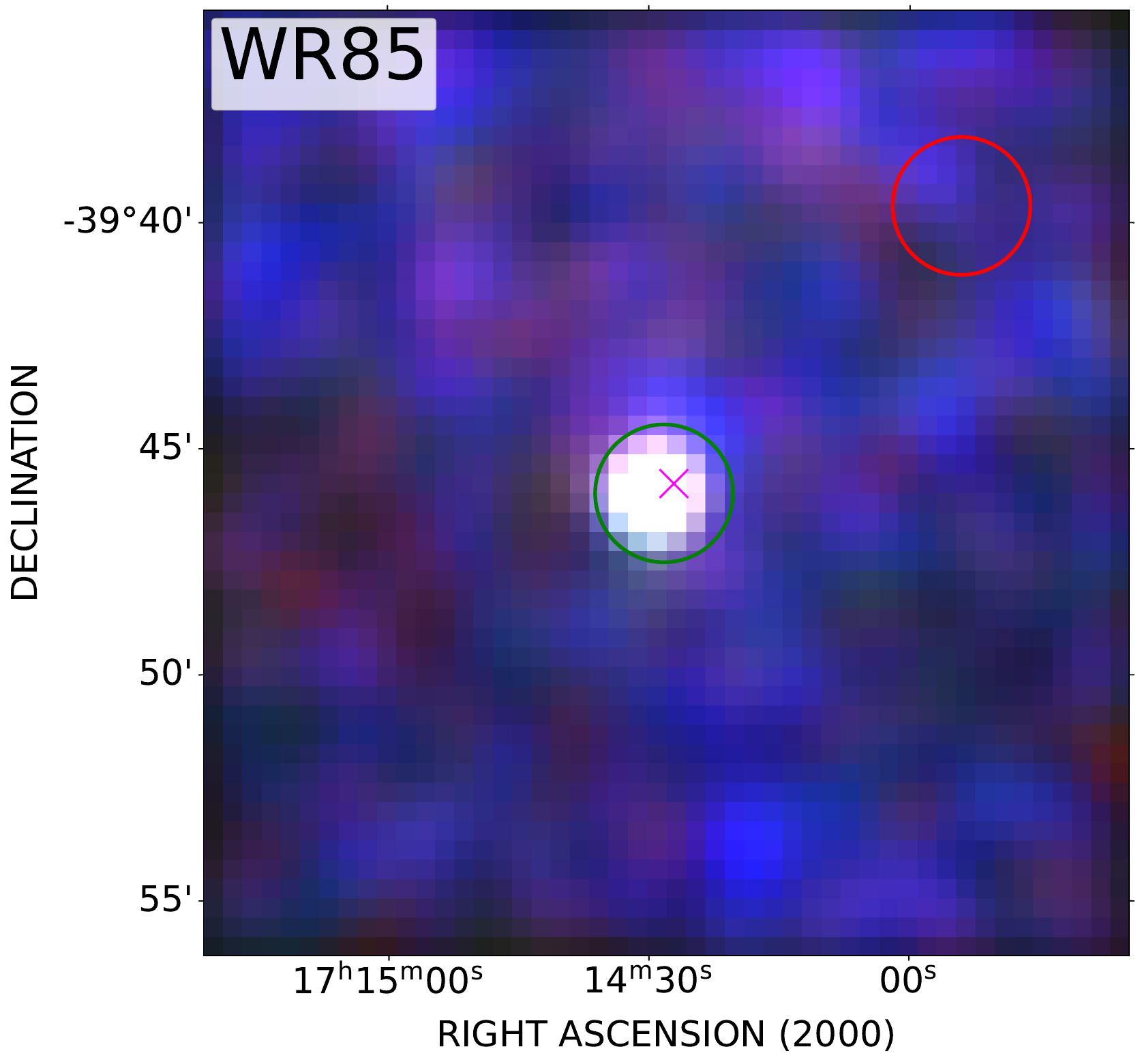}
    \includegraphics[scale=0.17]{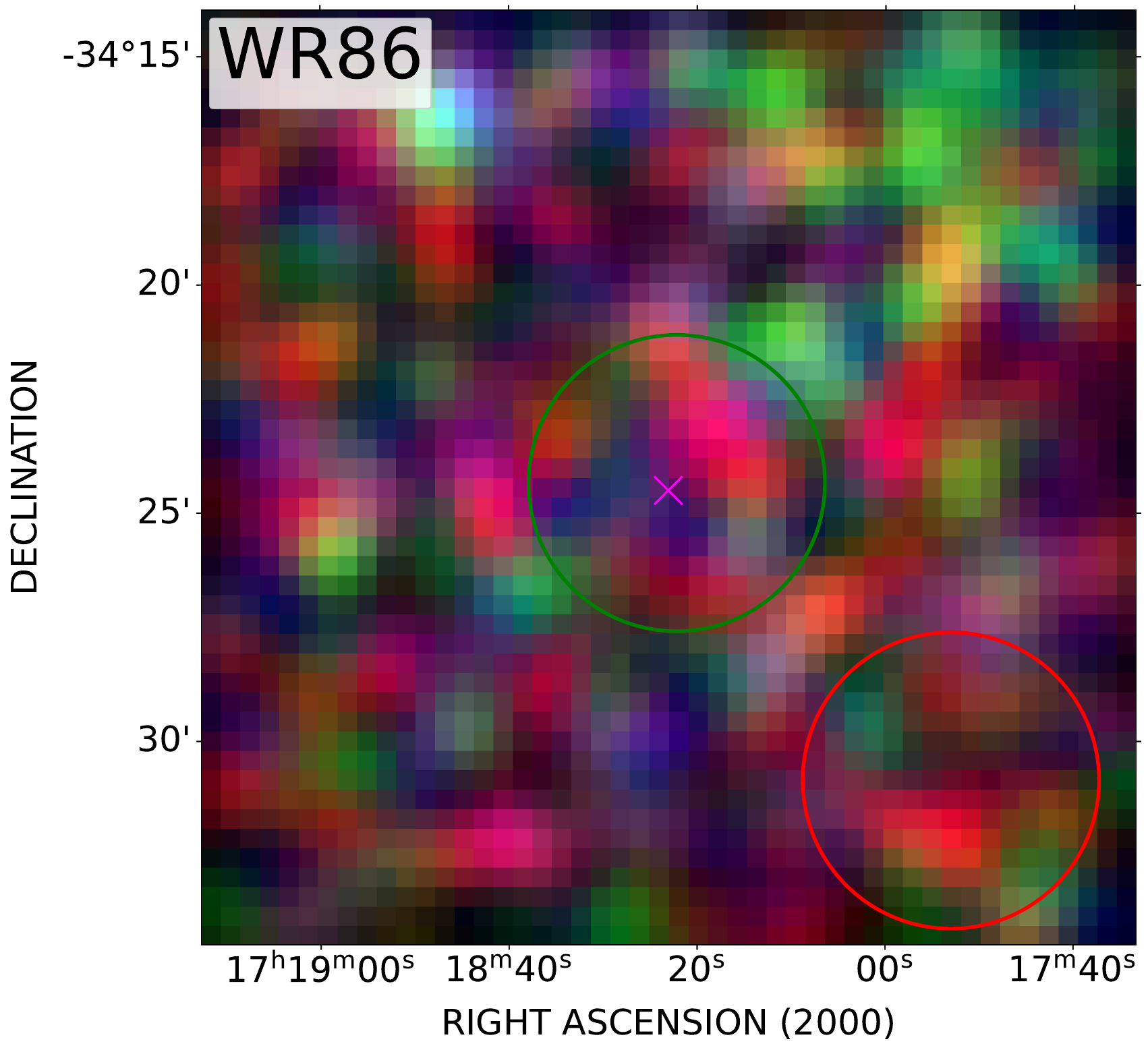}
    \includegraphics[scale=0.17]{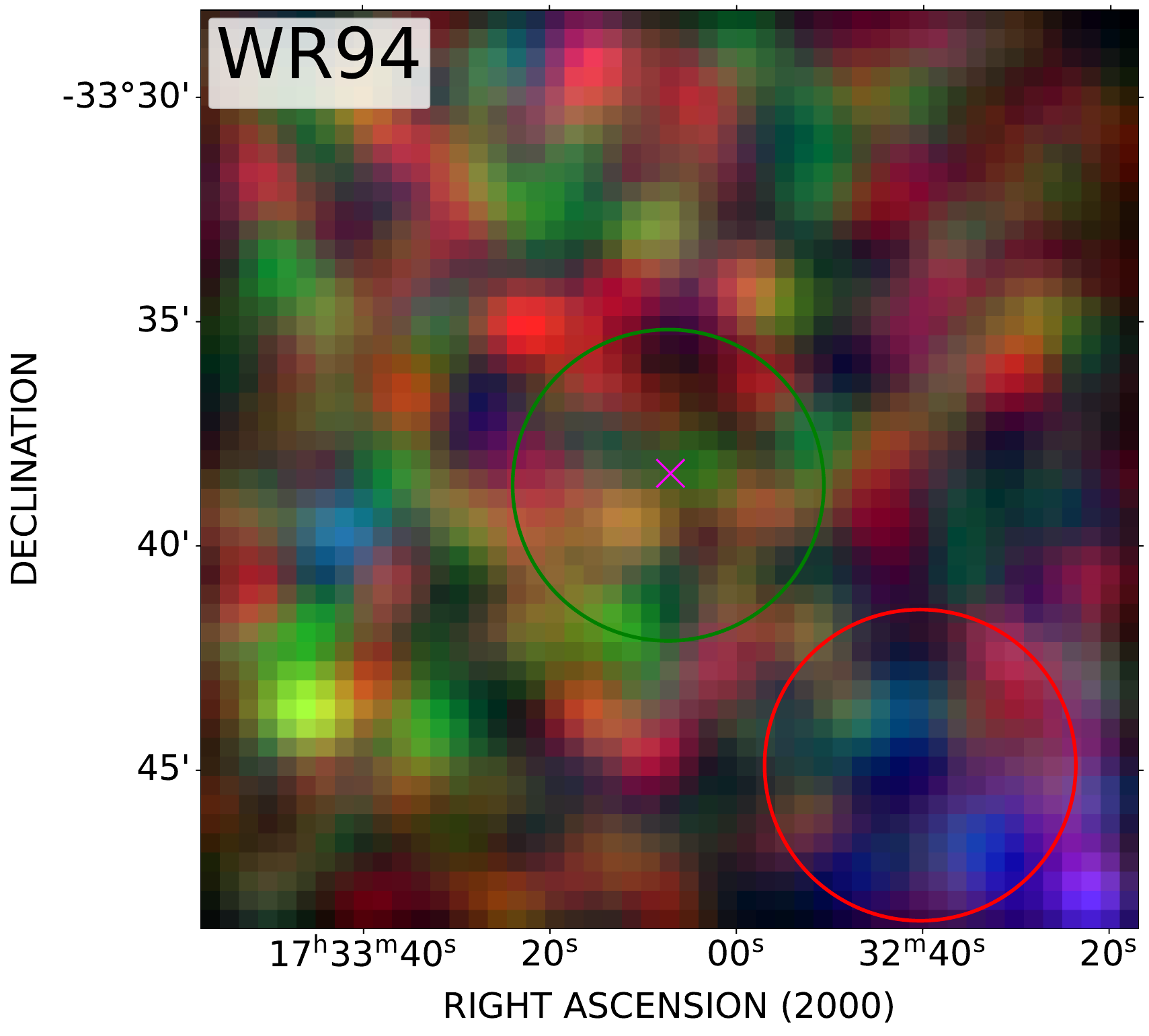}
    \includegraphics[scale=0.17]{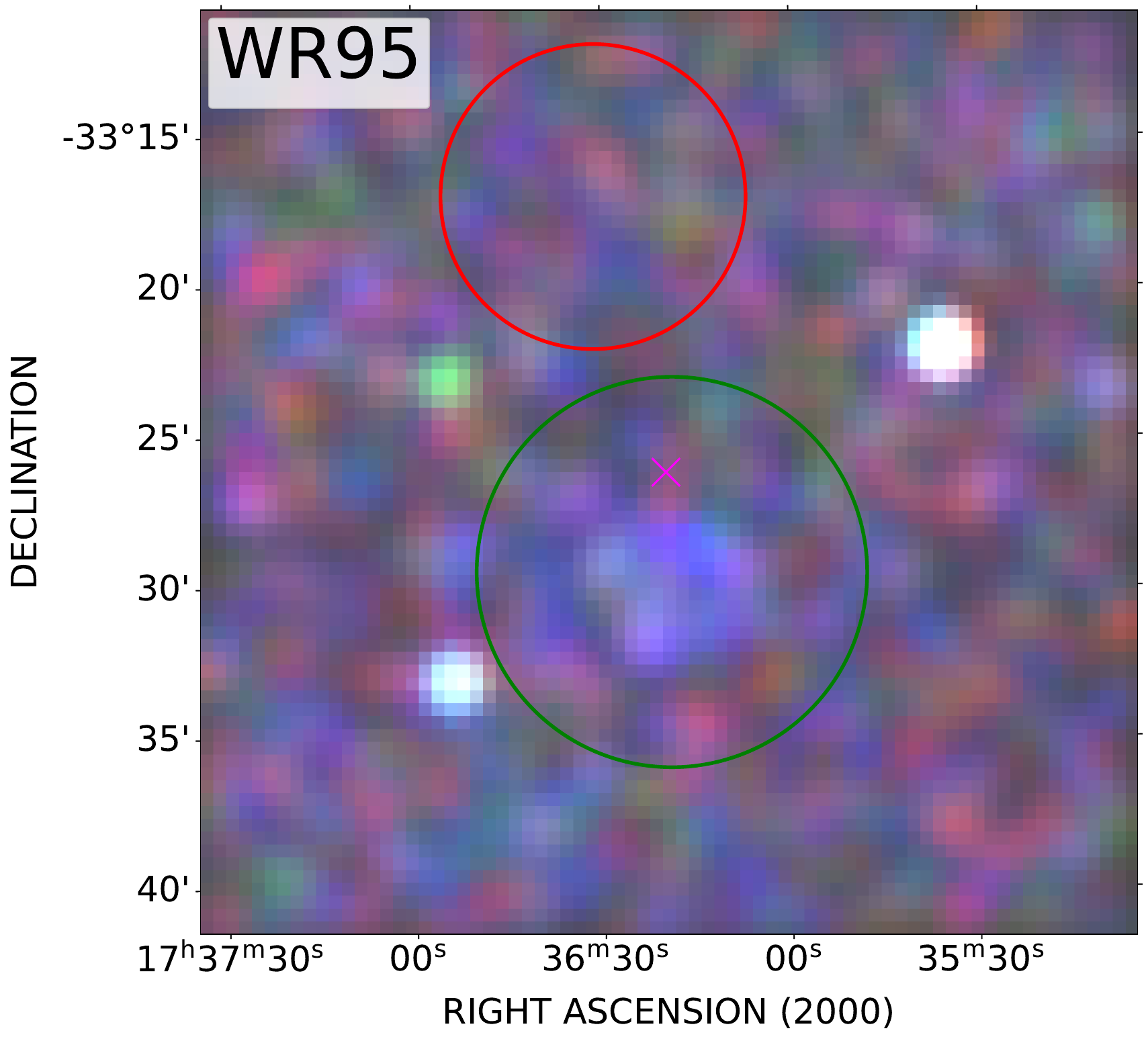}
    \includegraphics[scale=0.17]{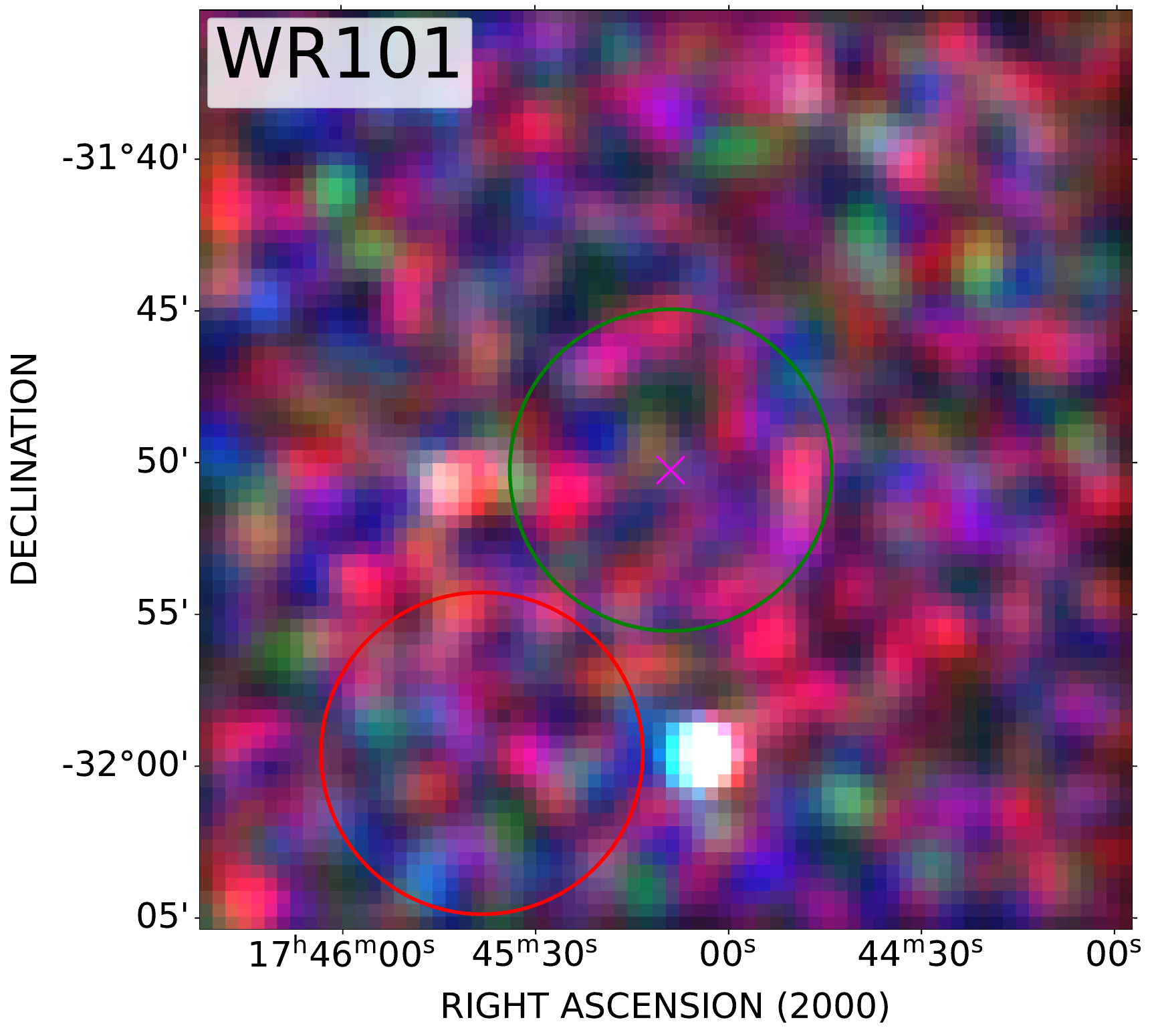}
    \caption{Exposure corrected images in X-rays (Red: 0.2-0.7 keV. Green: 0.7-1.1 keV. Blue: 1.1-10 keV) of the regions surrounding the WR stars listed in Table \ref{tab:Bubbles_Fluxes}. \textit{Top left}:  Corresponding image from WR40 to WR101. We excluded the images of WR68 whose emission is contaminated by surrounding unrelated diffuse emission. In green and red, the source and background extraction regions are shown, respectively, employed for the flux estimation (Table \ref{tab:Bubbles_Fluxes}). In yellow, we show the WISE contours only for the most representative bubbles. A magenta cross indicates the position of the WR star.
    \label{fig:WR_40_101}}
\end{figure*}

 Several details that have been observed for a number of bubbles are displayed in Figures \ref{fig:WR_7_38} and  \ref{fig:WR_40_101}, while the explicit background determination in the flux estimation allows us to reject many false positives. The most intriguing cases are still WR75 and WR95 and  from our images, some diffuse-like emission can be spotted around the position of the WR star. Looking at the spectrum of WR75 (Figure \ref{fig:Spectrum_WR75}), another thermal component (source model) is definitely needed below 1 keV, where  the contribution of the background is lower. Further investigations with \textit{XMM-Newton} could be very useful to confirm the presence of an eventual X-ray emitting bubble around this WR star. 

\begin{figure}
\includegraphics[scale=0.63]{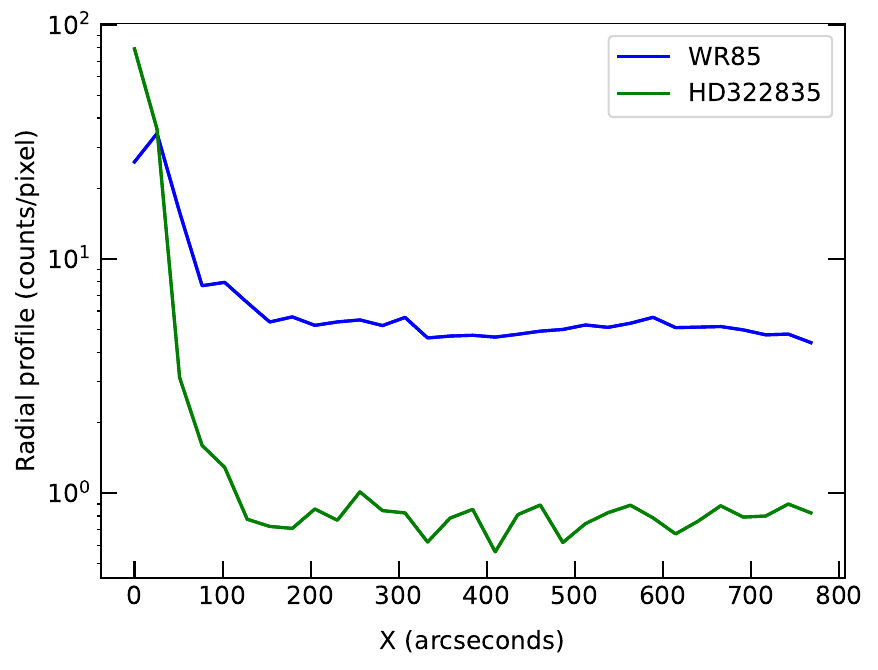}
\caption{Radial profile comparison between WR85 and a reference star (HD322835), expressed in physical units (1 pixel is equal to 25.6 arcseconds).
\label{fig:WR85_radialprofile}}
\end{figure}

One of the possible reasons for the lack of detections for such bubbles in X-rays is that in some cases, they could be quite compact and not spatially resolved by eROSITA. We tried a simple experiment to test this hypothesis, comparing the radial profile of WR85 with that of one nearby known star (HD322835), which is clearly a point source. We selected this particular WR star as candidate since in the \textit{WISE} image shown by \cite{Toala2015} its bubble-shaped emission is quite evident and in X-rays, it appears to be very compact (the radial profiles are presented in Figure \ref{fig:WR85_radialprofile}). WR85 is embedded in the diffuse emission of the supernova remnant RX J1713.7-3946, resulting in a higher background level; however, looking at the shape of the radial profile one can spot a small deviation from the profile of a point source. The superior angular resolution of \textit{Chandra} would be really helpful in ultimately understanding whether the source is extended in X-rays or not. 

Another explanation for non detecting most of these bubbles comes from a spectroscopic argument. Looking at the image of S308 (Figure \ref{fig:S308_WR6}) and its spectrum (Figure \ref{fig:Spectrum_2vapec_extraN}), the emission is highly peaked in very soft X-rays and our spectral fits retrieve very low values for the column density. Therefore, it is very likely that absorption effects should play an important role with increasing distance and this would explain why most of the bubbles remain undetected in X-rays. An interesting case is represented by WR22 and WR85, where the emission of the Wolf-Rayet star is quite clear; however, this is not necessary true for many of the other stars we studied, suggesting that absorption might play a significant role. We recall how we have masked the point source for the significance and flux estimation. As an additional indication, the relatively high galactic latitude of WR6 provides an explanation for the lack of absorption and the consequent bright emission observed in X-rays: the mechanism is clearly described in \cite{Meyer2021}. Checking this conclusion with the optical extinction database\footnote{\url{https://astro.acri-st.fr/gaia_dev/}} of \cite{Lallement2019},  at 1.5 kpc, we find A$_{V}=0.12$ in the direction of WR6. From using the relation of \cite{Predehl1995}, this optical extinction value is equivalent to a column density in X-rays of $\approx 10^{20}$ cm$^{-2}$, which is fully in accordance with our data.

On top of these physical arguments, we recall that the total exposure provided by eROSITA in scanning mode is quite short comparing to \textit{XMM-Newton} pointed observation employed to discover the other X-ray detected bubbles. This provides a possible explanation for why we do not have a significant detection for the bubbles surrounding WR7 and WR18.

\section{Conclusion \label{sec:Conclusion}}
In this work, we provide the first complete spectral characterization in X-rays of the wind-blown bubble S308 surrounding the WR star WR6, together with the first image in its entire extent. We find that collision equilibrium and non-equilibrium thermal models provide substantially similar fit statistic, suggesting new observations with much higher spectral resolution are needed to assess the real nature of these bubbles. Moreover, we also searched for new wind-blown bubbles around other WR stars and we find none are significant above the background. However, we were able to provide flux upper limits. As discussed in Section \ref{sec:Discussion}, the main reasons for these non detections are likely to be absorption and compactness of the nebulae. Nevertheless, the theoretical picture to explain why these bubbles have lower temperature and luminosities than expected remains unclear. Specifically, local absorption, instabilities, and clumps with low X-ray luminosity have been advocated as a viable explanation to explain such observational facts \citep{Toala2011}. However, a recent discussion presented by \cite{Dwarkadas2023} shows how the debate is still open, with momentum-driven solution and asymmetries  that could play a more significant role than considered before. New data with a much higher spatial and spectral resolution would be very helpful to finally clear the picture about these intriguing objects.

\begin{acknowledgements}
 We would like to thank the referee for the comments that helped to improve the paper. We thank Nik Szymanek \& Telescope Live for kindly providing the optical image shown in Figure \ref{fig:S308_optical_composite}, which can be found at \url{https://www.flickr.com/photos/36672102@N07/}. We also would like to thank all the eROSITA team for the helpful discussions and suggestions provided during the realization of the paper. FC acknowledges support from the Deutsche Forschungsgemeinschaft through the grant BE 1649/11-1 and from the International Max-Planck Research School on Astrophysics at the Ludwig-Maximilians University (IMPRS). 
 This work is based on data from eROSITA, the soft X-ray instrument aboard SRG, a joint Russian-German science mission supported by the Russian Space Agency (Roskosmos), in the interests of the Russian Academy of Sciences represented by its Space Research Institute (IKI), and the Deutsches Zentrum für Luft- und Raumfahrt (DLR). The SRG spacecraft was built by Lavochkin Association (NPOL) and its subcontractors, and is operated by NPOL with support from the Max Planck Institute for Extraterrestrial Physics (MPE). The development and
construction of the eROSITA X-ray instrument was led by MPE, with contributions from the Dr. Karl Remeis Observatory Bamberg \& ECAP (FAU Erlangen-Nuernberg), the University of Hamburg Observatory, the Leibniz Institute for Astrophysics Potsdam (AIP) and the Institute for Astronomy and Astrophysics of the University of Tübingen, with the support of DLR and the Max Planck Society. The Argelander Institute for Astronomy of the University of Bonn and the
Ludwig Maximilians Universität Munich also participated in the science preparation for eROSITA. The eROSITA data shown here were processed using the
eSASS software system developed by the German eROSITA consortium. This work makes use of the Astropy Python package\footnote{\url{https://www.astropy.org/}} \citep{2013A&A...558A..33A,2018AJ....156..123A}. A particular mention goes to the in-development coordinated package of Astropy for region handling called Regions\footnote{\url{https://github.com/astropy/regions}}. We acknowledge also the use of Python packages Matplotlib \citep{Hunter:2007}, Scipy \citep{2020SciPy-NMeth}, PyLaTex\footnote{\url{https://github.com/JelteF/PyLaTeX/}} and NumPy \citep{harris2020array}. 
\end{acknowledgements}

\bibliography{bibliography.bib}
\bibliographystyle{aasjournal}

\end{document}